\documentclass[sigconf,nonacm]{acmart}

\usepackage{subfigure}
\usepackage{siunitx}
\usepackage{enumitem}

\usepackage{algorithm}
\usepackage[noend]{algpseudocode}
\let\ReturnInline\Return
\renewcommand{\Return}{\State\ReturnInline}
\algrenewcommand\algorithmicrequire{$\rhd$}
\algrenewcommand\algorithmicensure{$\square$}

\AtBeginDocument{%
  \providecommand\BibTeX{{%
    \normalfont B\kern-0.5em{\scshape i\kern-0.25em b}\kern-0.8em\TeX}}}

\setcopyright{acmcopyright}
\copyrightyear{2018}
\acmYear{2018}
\acmDOI{XXXXXXX.XXXXXXX}

\acmConference[Conference acronym 'XX]{Make sure to enter the correct
  conference title from your rights confirmation emai}{June 03--05,
  2018}{Woodstock, NY}

\newcommand{\ignore}[1]{}

\begin{document}

\title[An Incrementally Expanding Approach for Updating PageRank on Dynamic Graphs]{An Incrementally Expanding Approach for \\Updating PageRank on Dynamic Graphs}


\author{Subhajit Sahu}
\email{subhajit.sahu@research.iiit.ac.in}
\affiliation{%
  \institution{IIIT Hyderabad}
  \streetaddress{Professor CR Rao Rd, Gachibowli}
  \city{Hyderabad}
  \state{Telangana}
  \country{India}
  \postcode{500032}
}


\settopmatter{printfolios=true}

\begin{abstract}
PageRank is a popular centrality metric that assigns importance to the vertices of a graph based on its neighbors and their score. Efficient parallel algorithms for updating PageRank on dynamic graphs is crucial for various applications, especially as dataset sizes have reached substantial scales. This technical report presents our Dynamic Frontier approach. Given a batch update consisting of edge insertions and deletions, it progressively identifies affected vertices that are likely to change their ranks with minimal overhead. On a server equipped with a 64-core AMD EPYC-7742 processor, our Dynamic Frontier PageRank outperforms Static, Naive-dynamic, and Dynamic Traversal PageRank by $7.8\times$, $2.9\times$, and $3.9\times$ respectively - on uniformly random batch updates of size $10^{-7}|E|$ to $10^{-3}|E|$. In addition, our approach improves performance at an average rate of $1.8\times$ for every doubling of threads.
\end{abstract}

\keywords{Parallel PageRank algorithm, Dynamic Frontier approach}


\maketitle

\section{Introduction}
\label{sec:introduction}
PageRank \cite{rank-page99} is an algorithm that measures the importance of nodes in a network by assigning numerical scores based on the structure of links. It finds applications in web page ranking, identifying misinformation, predicting traffic flow, and protein target identification. The increasing availability of vast amounts of data represented as graphs has led to a significant interest in parallel algorithms for computing PageRank \cite{rank-garg16, rank-nvgraph, rank-giri20, rank-sarma13}.\ignore{--- it has been implemented on multicore CPUs \cite{rank-garg16}, GPUs \cite{rank-nvgraph}, FPGAs \cite{rank-guoqiang20}, SpMV ASICs \cite{rank-sadi18}, CPU-GPU hybrids \cite{rank-giri20}, CPU-FPGA hybrids \cite{rank-li21}, and distributed systems \cite{rank-sarma13}.}

However, most real-world graph evolve with time. Here, frequent edge insertions and deletions make recomputing PageRank from scratch impractical, particularly for small, rapid changes. Existing strategies optimize by iterating from the prior snapshot's ranks, reducing the number of iterations needed for convergence. For further improvements, it is essential to recompute only the ranks of vertices likely to change. A prevalent approach involves identifying reachable vertices from the updated regions of the graph, and limiting processing to such vertices. However, if updates are randomly distributed, they often fall within dense graph regions, necessitating processing of a substantial portion of the graph.

To reduce computational effort, one can incrementally expand the set of affected vertices starting from the updated graph region, rather than processing all reachable vertices from the first iteration. Additionally, it is possible to skip processing a vertex's neighbors if the change in its rank is small and is expected to have minimal impact on the ranks of its neighboring vertices. This technical report introduces such an approach.

\subsection{Our Contributions}

This report introduces our Dynamic Frontier approach\footnote{\url{https://github.com/puzzlef/pagerank-openmp-dynamic}}, which, when given a batch update involving edge insertions and deletions, incrementally identifies affected vertices likely to undergo rank changes with minimal overhead. On a server equipped with a 64-core AMD EPYC-7742 processor, our Dynamic Frontier PageRank surpasses Static, Naive-dynamic, and Dynamic Traversal PageRank by $7.8\times$, $2.9\times$, and $3.9\times$ respectively, for uniformly random batch updates of size $10^{-7}|E|$ to $10^{-3}|E|$, where $|E|$ is the number of edges in the original graph. Additionally, our approach exhibits a performance improvement of $1.8\times$ for each doubling of threads.

\section{Related work}
\label{sec:related}
\ignore{Some of the early work on dynamic graph algorithms in the sequential setting include the seminal sparsification method of Eppstein et al. \cite{graph-eppstein97} and the bounded incremental computation idea of Ramalingam \cite{incr-ramalingam96}. The latter advocates measuring the work done as part of the update in proportion to the effect the update has on the computation.}

A number of approaches have been proposed for performing incremental computation (updating PageRank values in a dynamic / evolving graph) of approximate PageRank. Chien et al. \cite{rank-chien01} identify a tiny region of the graph near the updated vertices and model the remainder of the graph as a single vertex in a new, much smaller graph. PageRanks are computed for the small graph and then transferred to the original graph. Chen et al. \cite{chen2004local} propose a number of methods to estimate the PageRank score of a particular web page using only a small subgraph of the entire web, by expanding backwards from the target node following reverse hyperlinks. Bahmani et al. \cite{bahmani2010fast} analyze the efficiency of Monte Carlo methods for incremental computation of PageRank. Zhan et al. \cite{zhan2019fast} propose a Monte Carlo based algorithm for PageRank tracking on dynamic networks, by maintaining $R$ random walks starting from each node. Pashikanti et al. \cite{rank-pashikanti22} also follow a similar approach for updating PageRank scores on vertex and edge insertion/deletion.

A few approaches have been proposed for updating exact PageRank scores on dynamic graphs. Zhang \cite{rank-zhang17} presents a simple incremental Pagerank computation system for dynamic graphs on hybrid CPU and GPU platforms that incorporates the Update-Gather-Apply-Scatter (UGAS) computation model. A common approach used for Dynamic PageRank algorithm, given a small change to the input graph, is to find the affected region in the preprocessing step with Breadth-First Search (BFS) or Depth-First Search (DFS) traversal from the vertices connecting the edges that were inserted or deleted, and computing PageRanks only for that region \cite{rank-desikan05, kim2015incremental, rank-giri20, sahu2022dynamic}. This approach was originally proposed by Desikan et al. \cite{rank-desikan05}. Kim and Choi \cite{kim2015incremental} use this approach with an asynchronous implementation of PageRank. Giri et al. \cite{rank-giri20} use this approach with collaborative executions on muti-core CPUs and massively parallel GPUs. Sahu et al. \cite{sahu2022dynamic} use this approach on a Strongly Connected Component (SCC) based decomposition of the graph to limit the computation to SCCs that are reachable from updated vertices, on multi-core CPUs and GPUs (separately). Ohsaka et al. \cite{ohsaka2015efficient} propose an approach for locally updating PageRank using the Gauss-Southwell method, where the vertex with the greatest residual is updated first --- however, their algorithm is inherently sequential. 


Further, Bahmani et al. \cite{rank-bahmani12} propose an algorithm to selectively crawl a small portion of the web to provide an estimate of true PageRank of the graph at that moment, while Berberich et al. \cite{rank-berberich07} present a method to compute normalized PageRank scores that are robust to non-local changes in the graph. Their approaches are orthogonal to our \textit{Dynamic Frontier} approach which focuses on the computation of the PageRank vector itself, not on the process of crawling the web or maintaining normalized scores.

\section{Preliminaries}
\label{sec:preliminaries}
\subsection{PageRank algorithm}
\label{sec:pagerank}

The PageRank, $R[v]$, of a vertex $v \in V$ in the graph $G(V, E)$, represents its \textit{importance} and is based on the number of incoming links and their significance. Equation \ref{eq:pr} shows how to calculate the PageRank of a vertex $v$ in the graph $G$, with $V$ as the set of vertices ($n = |V|$), $E$ as the set of edges ($m = |E|$), $G.in(v)$ as the incoming neighbors of vertex $v$, $G.out(v)$ as the outgoing neighbors of vertex $v$, and $\alpha$ as the damping factor. Each vertex starts with an initial PageRank of $1/n$. The \textit{power-iteration} method updates these values iteratively until the change is rank values is within a specified tolerance $\tau$ value (indicating that convergence has been achieved).

\ignore{The \textit{random surfer model} is a conceptual framework for the PageRank algorithm, where a random surfer moves through the web by following the links on each page. The damping factor, $\alpha$, is the probability that the random surfer will continue to the next page along one of the links, instead of jumping to a random page on the web, and has a default value of $0.85$. The PageRank of each page can be seen as the long-term probability that the random surfer will visit that page, given that he starts on a random page and follows links according to the damping factor. The PageRank values can be calculated by finding the eigenvector of a transition matrix that represents the probabilities of moving from one page to another in the Markov Chain.}

Presence of dead ends is an issue that arises when computing the PageRank of a graph. A dead end is a vertex with no out-link, which forces the random surfer to jump to a random page on the web. Or equivalently, a dead end contributes its rank among all the vertices in the graph (including itself). This introduces a global teleport rank contribution that must be computed every iteration, and can be considered an overhead. We resolve this issue by adding self-loops to all the vertices in the graph \cite{rank-andersen07, rank-langville06}.

\begin{equation}
\label{eq:pr}
    R[v] = \alpha \times \sum_{u \in G.in(v)} \frac{R[u]}{|G.out(u)|} + \frac{1 - \alpha}{n}
\end{equation}

\subsection{Dynamic Graphs}
\label{sec:about-dynamic}

A dynamic graph can be viewed as a sequence of graphs, where $G^t(V^t, E^t)$ denotes the graph at time step $t$. The changes between graphs $G^{t-1}(V^{t-1}, E^{t-1})$ and $G^t(V^t, E^t)$ at consecutive time steps $t-1$ and $t$ can be denoted as a batch update $\Delta^t$ at time step $t$ which consists of a set of edge deletions $\Delta^{t-} = \{(u, v)\ |\ u, v \in V\} = E^{t-1} \setminus E^t$ and a set of edge insertions $\Delta^{t+} = \{(u, v)\ |\ u, v \in V\} = E^t \setminus E^{t-1}$.

\paragraph{Interleaving of graph update and computation:}

Changes to the graph arrive in a batched manner, with updating of the graph and execution of the desired algorithm being interleaved (i.e., there is only one writer upon the graph at a given point of time). In case it is desirable to update the graph while an algorithm is still running, a snapshot of the graph needs to be obtained, upon which the desired algorithm may be executed. See for example Aspen graph processing framework which significantly minimizes the cost of obtaining a read-only snapshot of the graph \cite{graph-dhulipala19}.

\subsection{Existing approaches for updating PageRank on Dynamic Graphs}

\subsubsection{Naive-dynamic approach}
\label{sec:about-naive}

This is a straightforward approach of updating ranks of vertices in dynamic networks. Here, one initializes the ranks of vertices with ranks obtained from previous snapshot of the graph and runs the PageRank algorithm on all vertices. Rankings obtained through this method will be at least as accurate as those obtained through the static algorithm.\ignore{Zhang et al. \cite{rank-zhang17} have explored the \textit{Naive-dynamic} approach in the hybrid CPU-GPU setting.}

\subsubsection{Dynamic Traversal approach}
\label{sec:about-traversal}

Originally proposed by Desikan et al. \cite{rank-desikan05}, here one skips processing of vertices that have no chance of their rank being updated as a result of the given batch update. For each edge deletion/insertion $(u, v)$ in the batch update, one marks all the vertices reachable from the vertex $u$ in the graph $G^{t-1}$ or the graph $G^t$ as affected (using DFS or BFS).\ignore{Giri et al. \cite{rank-giri20} have explored the \textit{Dynamic Traversal} approach in the hybrid CPU-GPU setting. On the other hand, Banerjee et al. \cite{rank-sahu22} have explored this approach in the CPU and GPU settings separately where they compute the ranks of vertices in topological order of strongly connected components (SCCs) to minimize unnecessary computation. They borrow this ordered processing of SCCs from the original static algorithm proposed by Garg et al. \cite{rank-garg16}.}

\section{Approach}
\label{sec:approach}
\subsection{Our Dynamic Frontier approach}
\label{sec:frontier}

If a batch update $\Delta^{t-} \cup \Delta^{t+}$ is small compared to the total number of edges $|E|$, then it is expected that the ranks of only a few vertices change. Our proposed Dynamic Frontier approach incorporates this aspect, and identifies affected vertices via an incremental process. This allows it to avoid unnecessary computation, since ranks of vertices far for the updated region of the graph cannot have a change in their ranks until the ranks of its immediate in-neighbors change. In addition, we avoid marking the neighbors of a vertex as affected, if the change in rank of the vertex is small enough and is likely to have minimal effect on the ranks of its neighbors.

\subsubsection{Explanation of the approach}
\label{sec:frontier-explanation}

Consider a batch update consisting of edge deletions $(u, v) \in \Delta^{t-}$ and insertions $(u, v) \in \Delta^{t+}$. We first initialize the rank of each vertex to that obtained in the previous snapshot of the graph.

\begin{figure*}[hbtp]
  \centering
  \subfigure[Initial graph]{
    \label{fig:about-frontier-01}
    \includegraphics[width=0.23\linewidth]{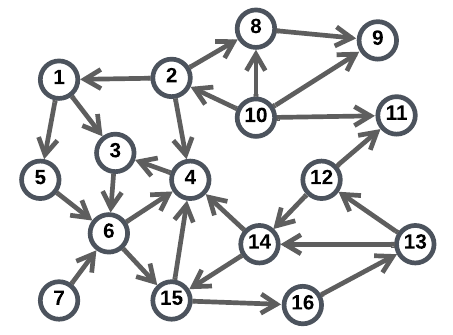}
  }
  \subfigure[Marking affected (initial)]{
    \label{fig:about-frontier-02}
    \includegraphics[width=0.23\linewidth]{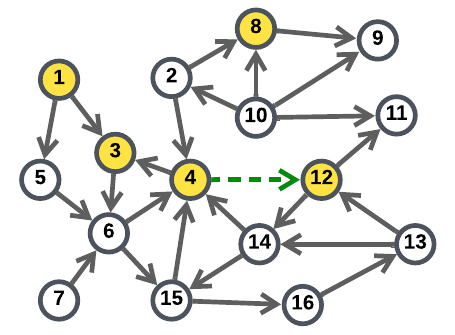}
  }
  \subfigure[After first iteration]{
    \label{fig:about-frontier-03}
    \includegraphics[width=0.23\linewidth]{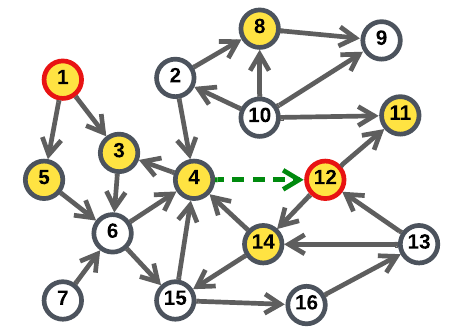}
  }
  \subfigure[After second iteration]{
    \label{fig:about-frontier-04}
    \includegraphics[width=0.23\linewidth]{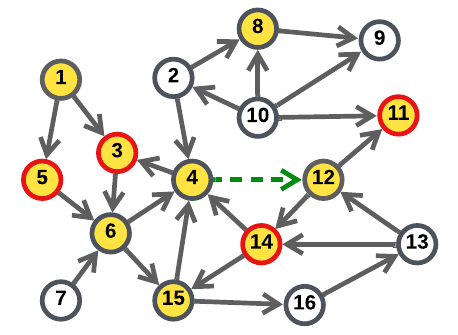}
  } \\[-2ex]
  \caption{Illustration of the \textit{Dynamic Frontier} approach through a specific example. The initial graph consists of $16$ vertices and $25$ edges. The graph is then updated with an edge insertion $(4, 12)$, and an edge deletion $(2, 1)$. Accordingly, the outgoing neighbors of vertices $4$ ($3$ and $12$) and $2$ ($1$, $4$, and $8$) are marked as affected (shown with yellow fill). When the ranks of these affected vertices are computed in the first iteration, it is found that change in rank of vertices $1$ and $12$ exceeds the frontier tolerance $\tau_f$ (shown with red border). Thus, outgoing neighbors of vertices $1$ ($3$ and $5$) and $12$ ($11$ and $14$) are also marked as affected. In the second iteration, the change in rank of vertices $3$, $5$, $11$, and $14$ is greater than $\tau_f$ --- thus their outgoing vertices are marked as affected. In the subsequent iteration, the ranks of affected vertices are again updated. If the change in rank of every vertex is within iteration tolerance $\tau$, the ranks of vertices have converged, and the algorithm terminates.}
  \label{fig:about-frontier}
\end{figure*}

\paragraph{Initial marking of affected vertex on edge deletion/insertion:}

For each edge deletion/insertion $(u, v)$, we initially mark the outgoing neighbors of the vertex $u$ in the previous $G^{t-1}$ and current graph snapshot $G^t$ as affected.

\paragraph{Incremental marking of affected vertices upon change in rank of a given vertex:}

Next, while performing PageRank computation, if the rank of any affected vertex $v$ changes in an iteration by an amount greater than the \textit{frontier tolerance} $\tau_f$, we mark its outgoing neighbors as affected. This process of marking vertices continues in every iteration.

\subsubsection{A simple example}

Figure \ref{fig:about-frontier} shows an example of the Dynamic Frontier approach. The initial graph, shown in Figure \ref{fig:about-frontier-01}, comprises $16$ vertices and $25$ edges. Subsequently, Figure \ref{fig:about-frontier-02} shows a batch update applied to the original graph involving the deletion of an edge from vertex $2$ to $1$ and the insertion of an edge from vertex $4$ to $12$. Following the batch update, we perform the initial step of the Dynamic Frontier approach, marking outgoing neighbors of $2$ and $4$ as affected, i.e., $1$, $3$, $4$, $8$, and $12$ are marked as affected (indicated with a yellow fill). Note that vertex $2$ is not affected as it is a source of the change while vertex $4$ being a neighbour of $2$ is marked as affected. Now, we are ready to execute the first iteration of PageRank algorithm.

During the first iteration (see Figure \ref{fig:about-frontier-03}), the ranks of affected vertices are updated. It is observed that the rank changes of vertices $1$ and $12$ surpass the frontier tolerance $\tau_f$ (highlighted with a red border). In response to this, we incrementally mark the outgoing neighbors of $1$ and $12$ as affected, i.e., vertices $3$, $5$, $11$, and $14$. 

During the second iteration (see Figure \ref{fig:about-frontier-04}), the ranks of affected vertices are again updated. Here, its is observed that the change in rank of vertices $3$, $5$, $11$, and $14$ is greater than frontier tolerance $\tau_f$. Thus, we mark the outgoing neighbors of $3$, $5$, $11$, and $14$ as affected, namely vertices $4$, $6$, and $15$. In the subsequent iteration, the ranks of affected vertices are again updated. If the change in rank of each vertex is within iteration tolerance $\tau$, the ranks of vertices have converged, and the algorithm terminates.

\subsection{Synchronous vs Asynchronous implementation}

In a synchronous implementation, separate input and output rank vectors are used, ensuring deterministic results for parallel algorithms through vector swapping at the end of each iteration. In contrast, an asynchronous implementation utilizes a single rank vector, potentially achieving faster convergence and eliminating memory copies for unaffected vertices in dynamic approaches\ignore{, but introduces non-deterministic results in parallel algorithms}.

\begin{figure*}[!hbt]
  \centering
  \subfigure{
    \label{fig:approach-async--mean}
    \includegraphics[width=0.48\linewidth]{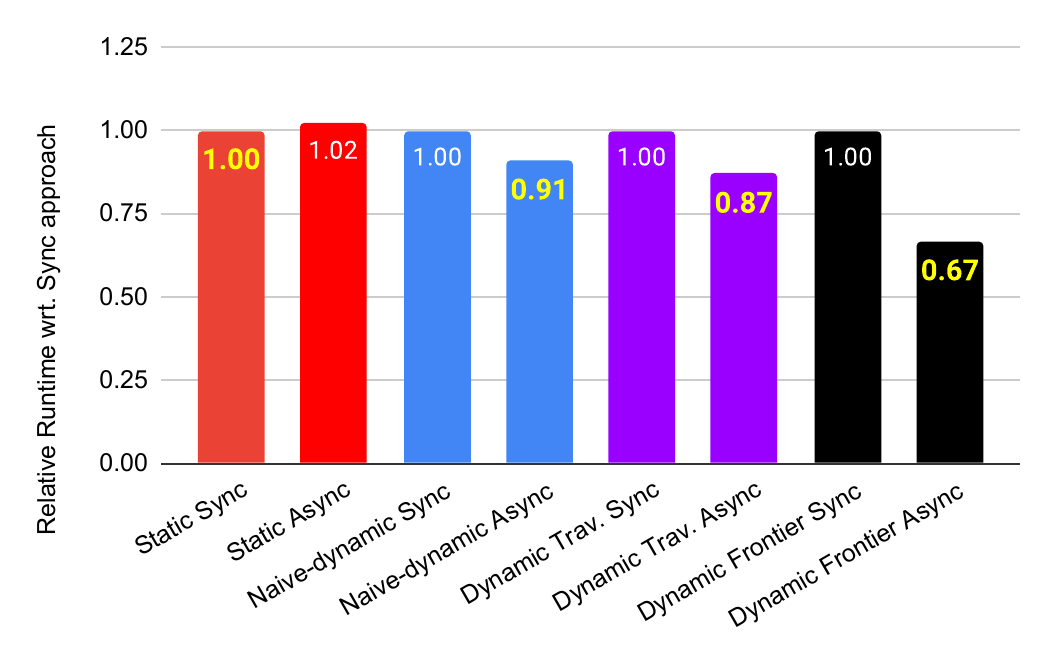}
  }
  \subfigure{
    \label{fig:approach-async--batch}
    \includegraphics[width=0.48\linewidth]{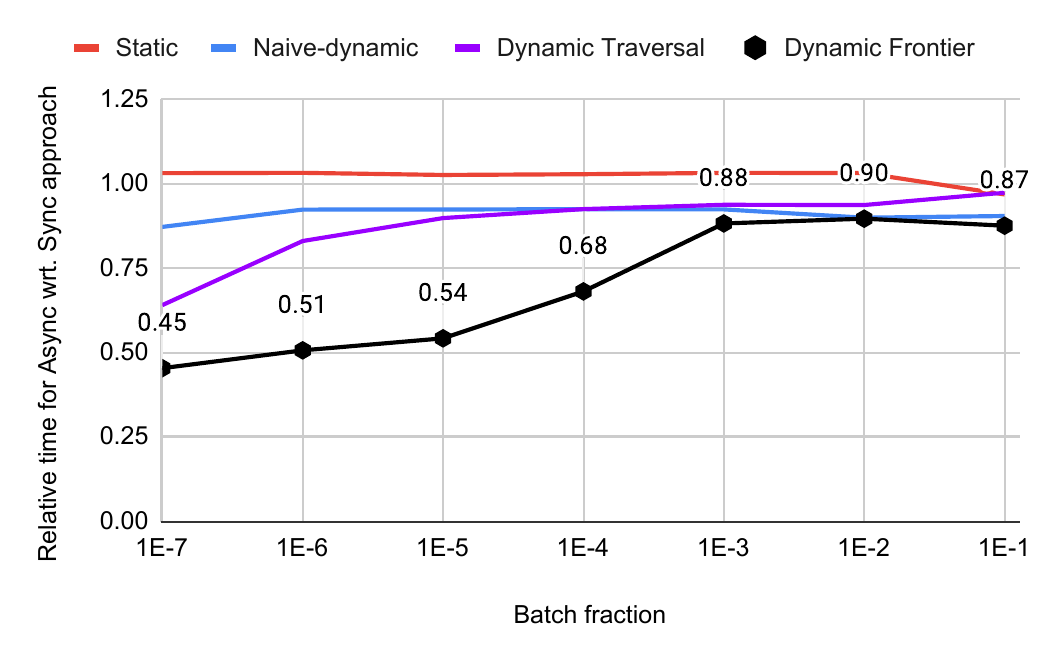}
  } \\[-2ex]
  \caption{Average Relative runtime with asynchronous implementations of \textit{Static}, \textit{Naive-dynamic}, \textit{Dynamic Traversal}, and \textit{Dynamic Frontier} approach compared to their respective synchronous implementations, on batch updates of size $10^{-7}|E|$ to $0.1|E|$ (right), and overall (left). The results indicate that asynchronous implementations are faster than synchronous ones, especially for smaller batch sizes. This is due to a somewhat faster convergence and the absence of copy overhead (for \textit{Dynamic Traversal} and \textit{Dynamic Frontier} approaches).}
  \label{fig:approach-async}
\end{figure*}

To assess synchronous and asynchronous implementations for Dynamic Frontier PageRank, both are tested on batch updates (purely edge insertions) ranging from $10^{-7}|E|$ to $0.1|E|$ for Static, Naive-dynamic, Dynamic Traversal, and Dynamic Frontier PageRank. Figure \ref{fig:approach-async} depicts the average relative runtime of asynchronous implementations compared to their synchronous counterparts. Based on the results, we use the asynchronous implementations of Naive-dynamic, Dynamic Traversal, and Dynamic Frontier PageRank --- as they are faster, especially for smaller batch sizes.\ignore{This is due to a somewhat faster convergence and the absence of copy overhead (for Dynamic Traversal and Dynamic Frontier approaches).}

\subsection{Determination of Frontier tolerance ($\tau_f$)}

We now measure a suitable value for frontier tolerance $\tau_f$ that allows us to minimize the number of vertices we process (after marking them as affected), while ensuring that we obtain ranks with the desired tolerance, i.e. we obtain ranks with no higher error than Static PageRank for the same tolerance setting. For this, we adjust frontier tolerance $\tau_f$ from $\tau$ to $\tau / 10^5$ and obtain ranks of vertices with the Dynamic Frontier approach on batch updates (consisting purely of edge insertions) of size $10^{-7}|E|$ to $0.1|E|$.

\begin{figure*}[!hbt]
  \centering
  \subfigure[Relative runtime with varying Frontier tolerance $\tau_f$]{
    \label{fig:adjust-frontier--runtime}
    \includegraphics[width=0.48\linewidth]{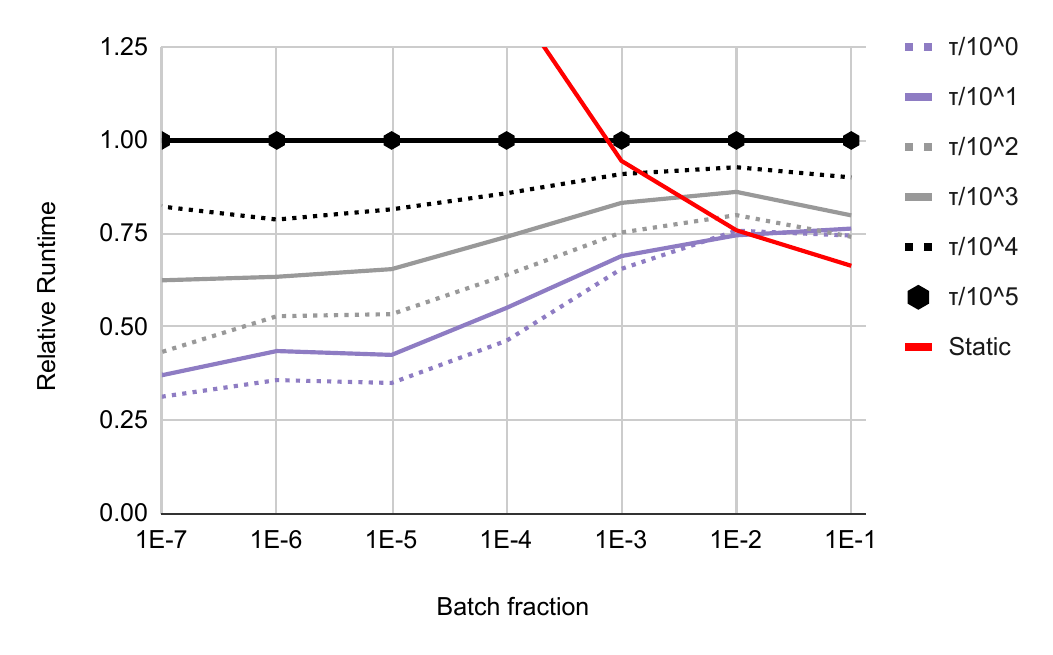}
  }
  \subfigure[Error in ranks obtained with varying Frontier tolerance $\tau_f$]{
    \label{fig:adjust-frontier--error}
    \includegraphics[width=0.48\linewidth]{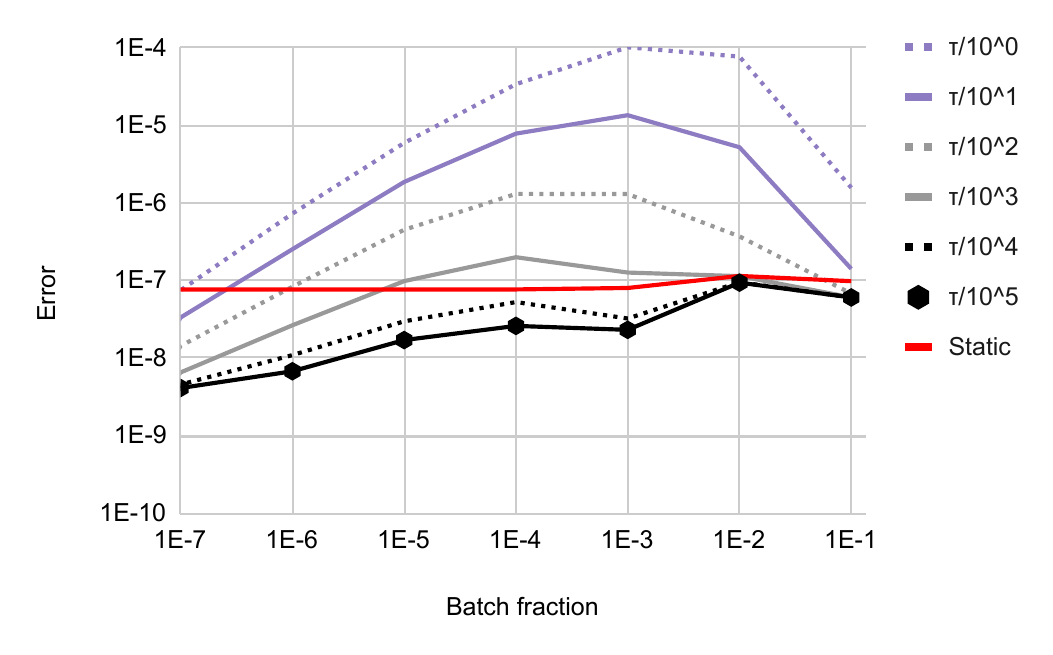}
  } \\[-2ex]
  \caption{Average Relative runtime and Error in ranks obtained (with respect to ranks obtained with Reference Static PageRank) using \textit{Dynamic Frontier} approach, with frontier tolerance $\tau_f$ varying from $\tau$ to $\tau / 10^5$, on batch updates of size $10^{-7}|E|$ to $0.1|E|$. The figures indicate that increasing $\tau_f$ reduces runtime, but also increases the error. A Frontier tolerance $\tau_f$ of $\tau/10^4$ and $\tau/10^5$ obtain ranks with error lower than \textit{Static} PageRank, and are thus acceptable (we choose $\tau_f = \tau/10^5$ to be on the safe side).}
  \label{fig:adjust-frontier}
\end{figure*}

Figure \ref{fig:adjust-frontier} illustrates the average relative runtime and rank error (in comparison to ranks obtained with reference Static PageRank) using the Dynamic Frontier approach. The figure suggests that as $\tau_f$ increases, runtime decreases, but it is accompanied by an increase in error. A frontier tolerance $\tau_f$ set at $\tau/10^4$ or $\tau/10^5$ yields ranks with lower error than Static PageRank, making them acceptable for uniformly random batch updates. To err on the side of caution, we opt for a frontier tolerance of $\tau_f = \tau/10^5$.

\begin{algorithm}[!hbt]
\caption{Our parallel Dynamic Frontier PageRank.}
\label{alg:frontier}
\begin{algorithmic}[1]
\Require{$G^{t-1}, G^t$: Previous, current input graph}
\Require{$\Delta^{t-}, \Delta^{t+}$: Edge deletions and insertions (input)}
\Require{$R^{t-1}$: Previous rank vector}
\Ensure{$R$: Current rank vector}
\Ensure{$\Delta r$: Change in rank of a vertex}
\Ensure{$\Delta R$: $L\infty$-norm between previous and current ranks}
\Ensure{$\tau, \tau_f$: Iteration, frontier tolerance}
\Ensure{$\alpha$: Damping factor}

\Statex

\Function{dynamicFrontier}{$G^{t-1}, G^t, \Delta^{t-}, \Delta^{t+}, R^{t-1}$}
  \State $R \gets R^{t-1}$
  \State $\rhd$ Mark initial affected
  \ForAll{$(u, v) \in \Delta^{t-} \cup \Delta^{t+} \textbf{in parallel}$} \label{alg:frontier--mark-begin}
    \ForAll{$v' \in (G^{t-1} \cup G^t).out(u)$}
    \State Mark $v'$ as affected
    \EndFor
  \EndFor \label{alg:frontier--mark-end}
  \ForAll{$i \in [0 .. MAX\_ITERATIONS)$} \label{alg:frontier--compute-begin}
    \State $\Delta R \gets 0$
    \ForAll{affected $v \in V^t$ \textbf{in parallel}}
      \State $r \gets (1 - \alpha)/|V^t|$
      \ForAll{$u \in G^t.in(v)$}
        \State $r \gets r + \alpha * R[u] / |G^t.out(u)|$
      \EndFor
      \State $\Delta r \gets |r - R[v]|$ \textbf{;} $R[v] \gets r$
      \State $\Delta R \gets max(\Delta R, \Delta r)$
      \State $\rhd$ Is rank change $>$ frontier tolerance?
      \If{$\Delta r > \tau_f$} \label{alg:frontier--remark-begin}
        \ForAll{$v' \in G^t.out(v)$}
          \State Mark $v'$ as affected
        \EndFor
      \EndIf \label{alg:frontier--remark-end}
    \EndFor
    \State $\rhd$ Ranks converged?
    \If{$\Delta R \le \tau$} \textbf{break}
    \EndIf
  \EndFor \label{alg:frontier--compute-end}
  \State \ReturnInline{$R$} \label{alg:frontier--return}
\EndFunction
\end{algorithmic}
\end{algorithm}



\subsection{Our Dynamic Frontier PageRank implementation}

Algorithm \ref{alg:frontier} shows our implementation of Dynamic Frontier PageRank, which is designed to compute the PageRank of vertices in a graph while efficiently handling dynamic changes in the graph structure over time. The algorithm takes as input the previous and current versions of the graph, edge deletions and insertions in the batch update, and the previous rank vector.

It begins by marking the initially affected vertices based on the edge deletions $\Delta^{t-}$ and insertions $\Delta^{t+}$ in parallel (lines \ref{alg:frontier--mark-begin}-\ref{alg:frontier--mark-end}). It then enters an iterative computation phase (lines \ref{alg:frontier--compute-begin}-\ref{alg:frontier--compute-end}), where it updates the rank of each affected vertex. The PageRank computation is performed in parallel for each affected vertex $v$, considering the incoming edges $G^t.in(v)$. The algorithm checks whether the change in rank $\Delta r$ exceeds the frontier tolerance $\tau_f$, and marks its out-neighbor vertices as affected if so. The iteration continues until either the net change in ranks $\Delta R$ (which is equal to the $L\infty$-norm between the previous and the current ranks) falls below the iteration tolerance $\tau$, or a maximum number of iterations is reached $MAX\_ITERATIONS$. In line \ref{alg:frontier--return}, the final rank vector $R$ is returned.


\section{Evaluation}
\label{sec:evaluation}
\subsection{Experimental Setup}
\label{sec:setup}

\subsubsection{System used}

We conduct experiments on a system equipped with an AMD EPYC-7742 processor, with $64$ cores and operating at a frequency of $2.25$ GHz. Each core has a $4$ MB L1 cache, a $32$ MB L2 cache, and shares a $256$ MB L3 cache. The server is configured with $512$ GB of DDR4 system memory and operates on Ubuntu $20.04$.

\subsubsection{Configuration}

We employ 32-bit integers for vertex ids and 64-bit floating-point numbers for vertex rankings. To denote affected vertices, an 8-bit integer vector is utilized. The rank computation utilizes OpenMP's \textit{dynamic schedule} with a chunk size of $2048$, facilitating dynamic workload balancing among threads. We use a damping factor of $\alpha = 0.85$ \cite{rank-langville06}, an iteration tolerance of $\tau = 10^{-10}$ using the $L_\infty$-norm \cite{rank-dubey22, rank-plimpton11}, and limit the maximum number of iterations (\texttt{MAX\_ITERATIONS}) to $500$ \cite{nvgraph}. We run all experiments with $64$ threads to match the number of cores available on the system (unless specified otherwise). Compilation is performed using GCC $9.4$ and OpenMP $5.0$.

\subsubsection{Dataset}

We use four graph classes sourced from the \textit{SuiteSparse Matrix Collection} \cite{suite19}, as detailed in Table \ref{tab:dataset}. The number of vertices in these graphs range from $3.07$ million to $214$ million, with edge counts spanning from $37.4$ million to $1.98$ billion. To address the impact of dead ends (vertices lacking out-links), a global teleport rank computation is needed in each iteration. We mitigate this overhead by adding self-loops to all vertices in the graph \cite{rank-andersen07, rank-langville06}.

\begin{table}[hbtp]
  \centering
  \caption{List of 12 graphs obtained from the SuiteSparse Matrix Collection \cite{suite19} (directed graphs are marked with $*$). Here, $|V|$ is the number of vertices, $|E|$ is the number of edges (after adding self-loops), and $D_{avg}$ is the average degree.\ignore{, and $\Gamma_G$ is the Gini coefficient of PageRank distribution. In the table, B refers to a billion, M refers to a million and K refers a thousand.}}
  \label{tab:dataset}
  \begin{tabular}{|c||c|c|c|c|}
    \toprule
    \textbf{Graph} &
    \textbf{\textbf{$|V|$}} &
    \textbf{\textbf{$|E|$}} &
    \textbf{\textbf{$D_{avg}$}} \\
    \midrule
    \multicolumn{4}{|c|}{\textbf{Web Graphs (LAW)}} \\ \hline
    indochina-2004$^*$ & 7.41M & 199M & 26.8 \\ \hline  
    arabic-2005$^*$ & 22.7M & 654M & 28.8 \\ \hline  
    uk-2005$^*$ & 39.5M & 961M & 24.3 \\ \hline  
    webbase-2001$^*$ & 118M & 1.11B & 9.4 \\ \hline  
    it-2004$^*$ & 41.3M & 1.18B & 28.5 \\ \hline  
    sk-2005$^*$ & 50.6M & 1.98B & 39.1 \\ \hline  
    \multicolumn{4}{|c|}{\textbf{Social Networks (SNAP)}} \\ \hline
    com-LiveJournal & 4.00M & 73.4M & 18.3 \\ \hline  
    com-Orkut & 3.07M & 237M & 77.3 \\ \hline  
    \multicolumn{4}{|c|}{\textbf{Road Networks (DIMACS10)}} \\ \hline
    asia\_osm & 12.0M & 37.4M & 3.1 \\ \hline  
    europe\_osm & 50.9M & 159M & 3.1 \\ \hline  
    \multicolumn{4}{|c|}{\textbf{Protein k-mer Graphs (GenBank)}} \\ \hline
    kmer\_A2a & 171M & 531M & 3.1 \\ \hline  
    kmer\_V1r & 214M & 679M & 3.2 \\ \hline  
  \bottomrule
  \end{tabular}
\end{table}

\subsubsection{Batch Generation}
\label{sec:batch-generation}

For each base (static) graph from the dataset, we generate a random batch update, consisting of purely edge insertions, purely edge deletions, or an $80\% : 20\%$ mix of edge insertions and deletions to mimic realistic batch updates. The set of edges for insertion is prepared by selecting vertex pairs with equal probability. To construct the set of edge deletions, we delete each existing edge with a uniform probability. For simplicity, we ensure that no new vertices are added to or removed from the graph. The batch size is measured as a fraction of edges in the original graph, and is varied from $10^{-7}$ to $0.1$ (i.e., $10^{-7}|E|$ to $0.1|E|$), with multiple batches generated for each size (for averaging). Along with each batch update, self-loops are added to all vertices.

\subsubsection{Measurement}
\label{sec:measurement}

We measure the time taken by each approach on the updated graph entirely, including any preprocessing costs and convergence detection time, while excluding time dedicated to memory allocation and deallocation. The mean time for a specific method at a given batch size is calculated as the geometric mean across various input graphs. Consequently, the average speedup is determined as the ratio of these mean times. Additionally, we gauge the error/accuracy of a given approach by assessing the $L1$-norm \cite{ohsaka2015efficient} of the ranks in comparison to ranks obtained from a reference Static PageRank run on the updated graph with an extremely low iteration tolerance of $\tau = 10^{-100}$ (limited to $500$ iterations).

\begin{figure*}[hbtp]
  \centering
  \subfigure[Overall result]{
    \label{fig:insertions-runtime--mean}
    \includegraphics[width=0.38\linewidth]{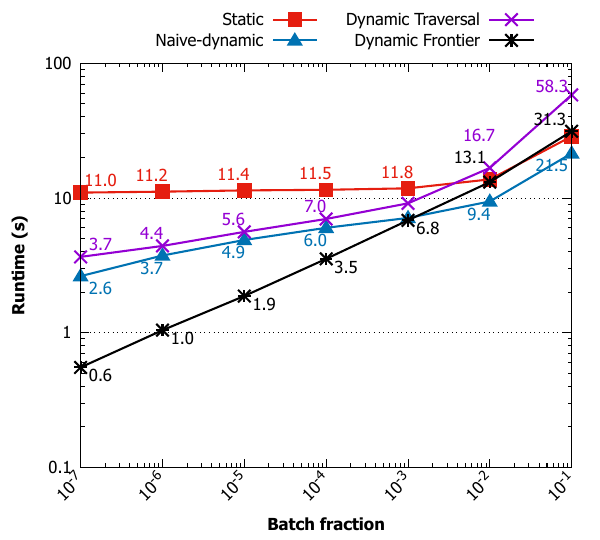}
  }
  \subfigure[Results on each graph]{
    \label{fig:insertions-runtime--all}
    \includegraphics[width=0.58\linewidth]{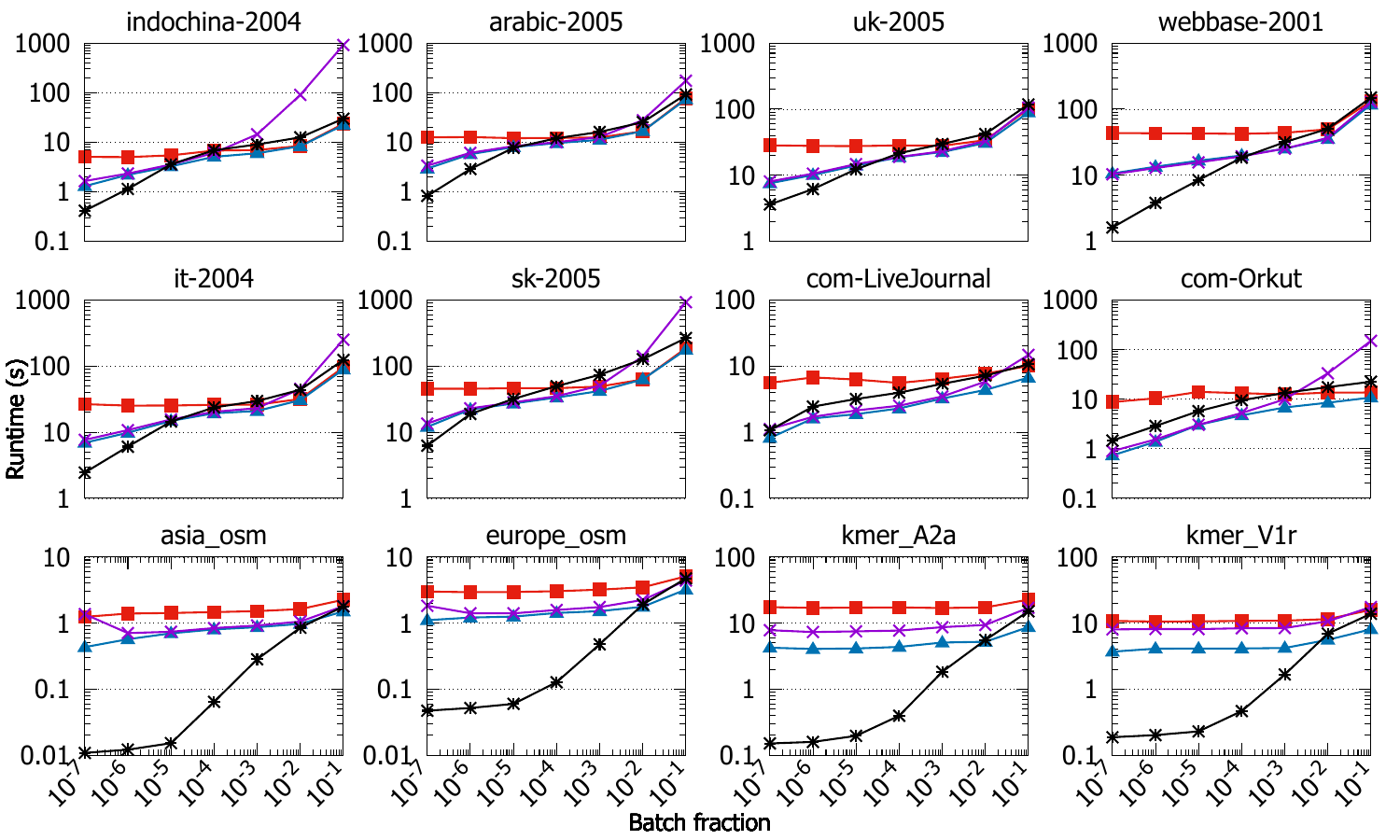}
  } \\[-1ex]
  \caption{Runtime (logarithmic scale) for \textit{Static}, \textit{Naive-dynamic}, \textit{Dynamic Traversal}, and \textit{Dynamic Frontier} PageRank with batch updates exclusively comprising edge insertions, ranging from $10^{-7} |E|$ to $0.1 |E|$ in multiples of $10$ (logarithmic scale). The right figure details the runtime of each approach for individual graphs in the dataset, while the left figure displays overall runtimes --- using geometric mean for consistent scaling across graphs.}
  \label{fig:insertions-runtime}
\end{figure*}

\begin{figure*}[hbtp]
  \centering
  \subfigure[Overall result]{
    \label{fig:insertions-speedup--mean}
    \includegraphics[width=0.38\linewidth]{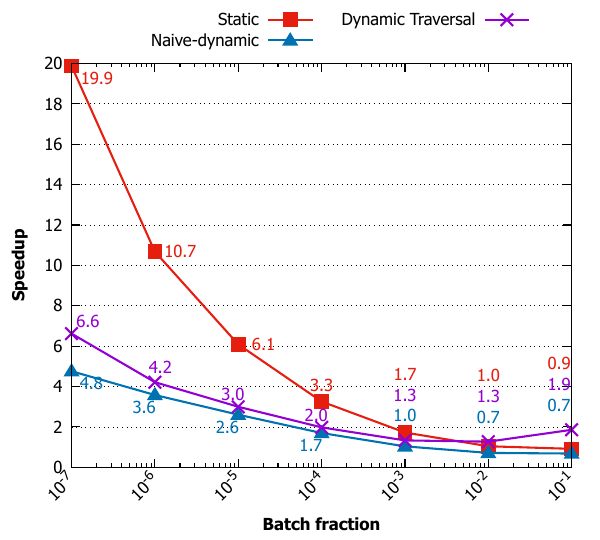}
  }
  \subfigure[Results on each graph]{
    \label{fig:insertions-speedup--all}
    \includegraphics[width=0.58\linewidth]{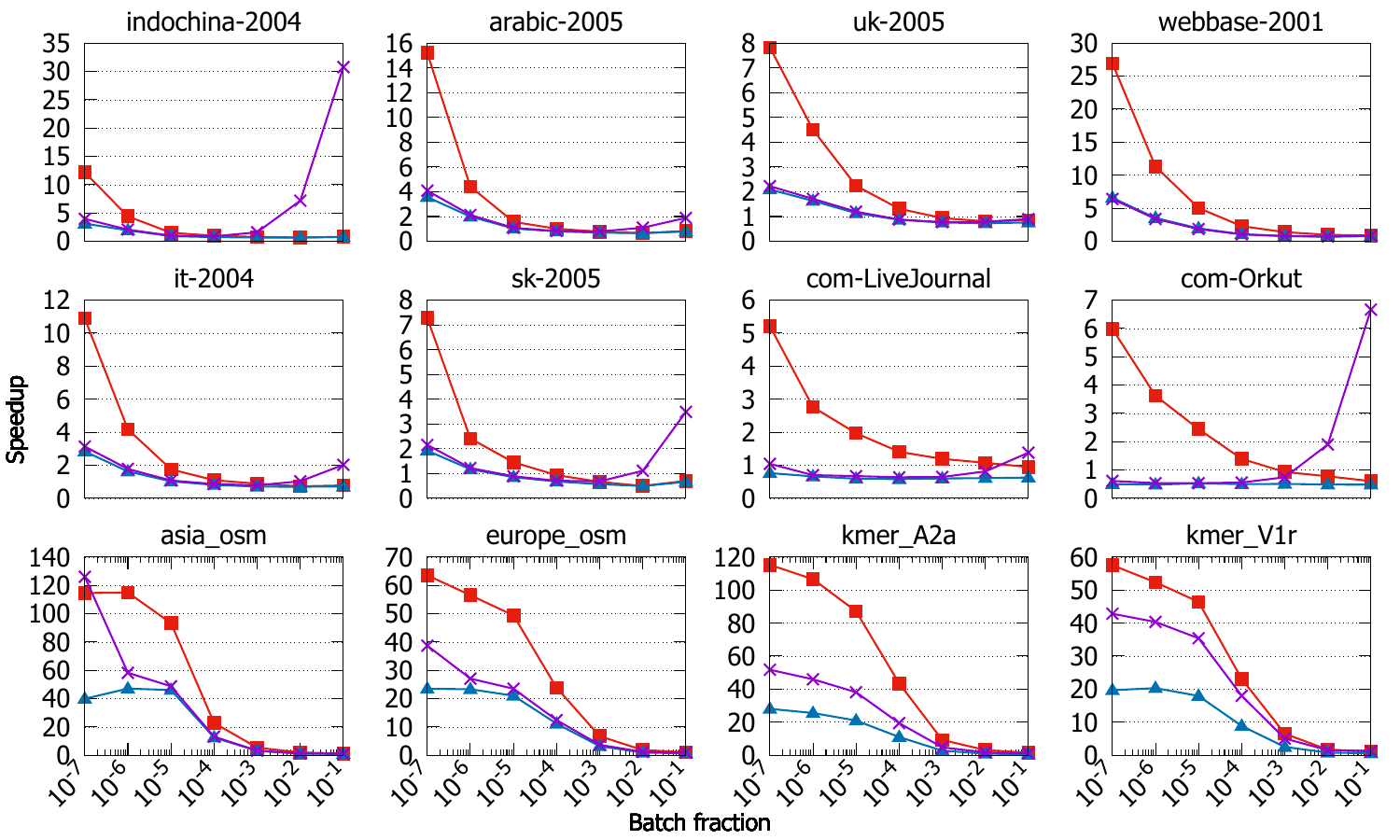}
  } \\[-1ex]
  \caption{Speedup of \textit{Dynamic Frontier} PageRank with respect to \textit{Static}, \textit{Naive-dynamic}, and \textit{Dynamic Traversal} PageRank, on batch updates consisting solely of edge insertions ranging from $10^{-7} |E|$ to $0.1 |E|$ (logarithmic scale). The right figure depicts the speedup of \textit{Dynamic Frontier} PageRank in relation to each approach for individual graphs in the dataset, while the left figure highlights the overall speedup.}
  \label{fig:insertions-speedup}
\end{figure*}

\begin{figure*}[hbtp]
  \centering
  \subfigure[Overall result]{
    \label{fig:insertions-error--mean}
    \includegraphics[width=0.38\linewidth]{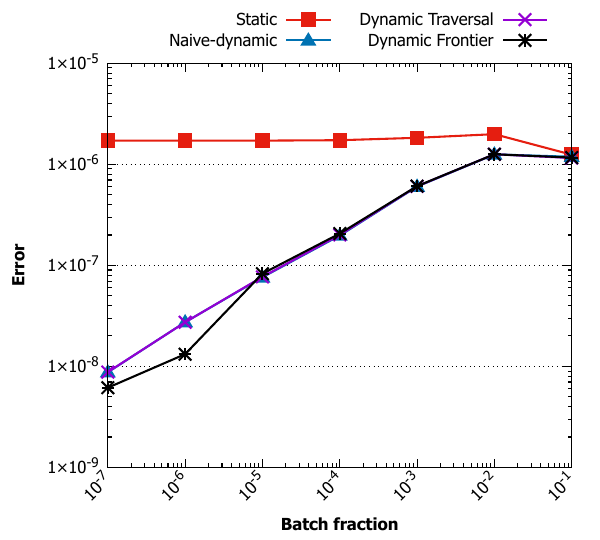}
  }
  \subfigure[Results on each graph]{
    \label{fig:insertions-error--all}
    \includegraphics[width=0.58\linewidth]{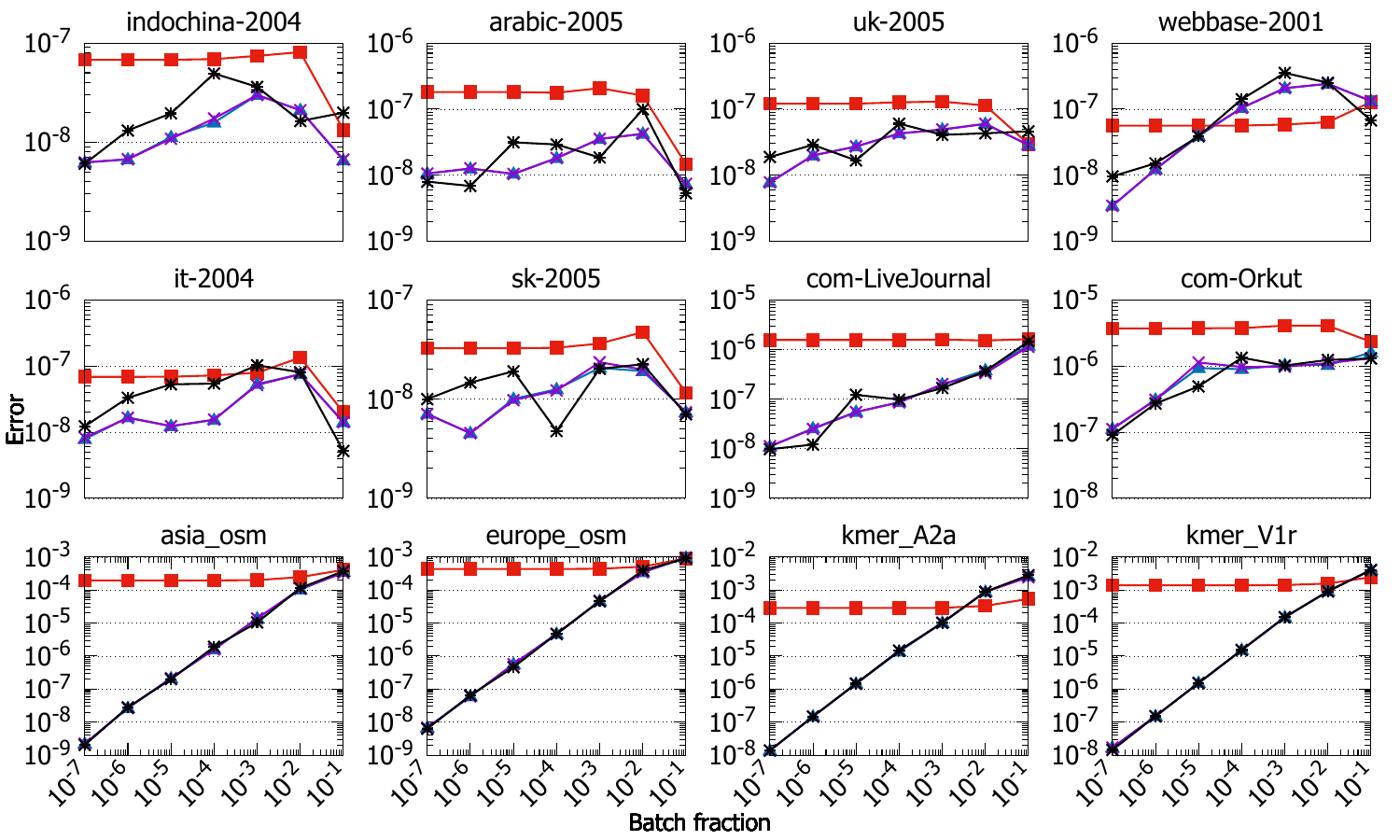}
  } \\[-1ex]
  \caption{Error analysis comparing \textit{Static}, \textit{Naive-dynamic}, \textit{Dynamic Traversal}, and \textit{Dynamic Frontier} PageRank with a Reference Static PageRank (with a tolerance $\tau$ of $10^{-100}$ and limited to $500$ iterations) using $L1$-norm. Batch updates involve edge insertions ranging from $10^{-7} |E|$ to $0.1 |E|$ (logarithmic scale). The right figure illustrates the error specific to each approach for individual graphs in the dataset, while the left figure presents overall errors using the geometric mean for consistent scaling across graphs.}
  \label{fig:insertions-error}
\end{figure*}

\begin{figure*}[hbtp]
  \centering
  \subfigure[Overall result]{
    \label{fig:deletions-runtime--mean}
    \includegraphics[width=0.38\linewidth]{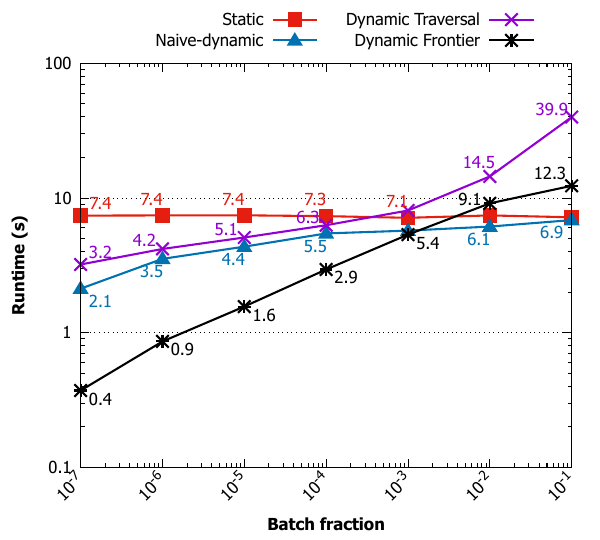}
  }
  \subfigure[Results on each graph]{
    \label{fig:deletions-runtime--all}
    \includegraphics[width=0.58\linewidth]{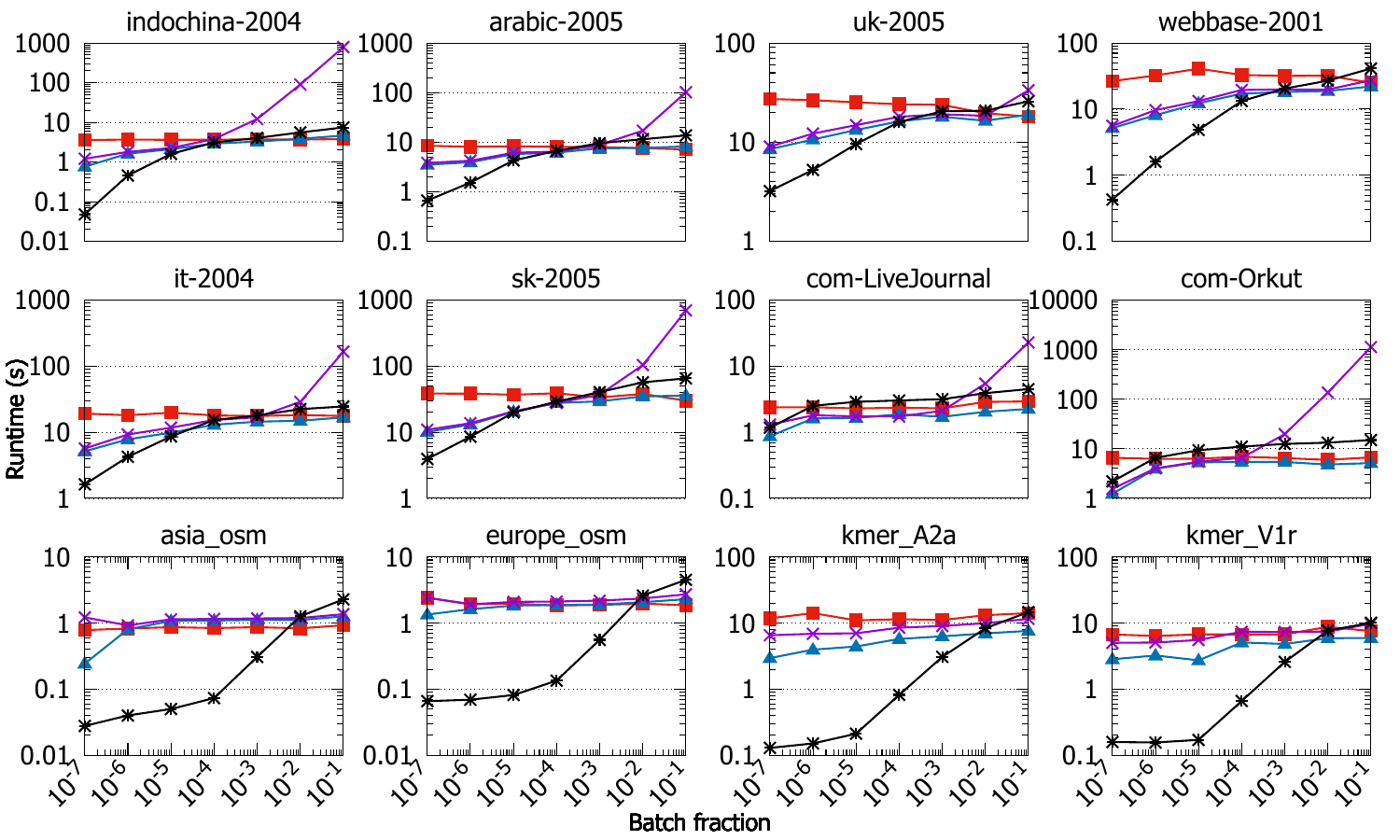}
  } \\[-1ex]
  \caption{Runtime (logarithmic scale) of \textit{Static}, \textit{Naive-dynamic}, \textit{Dynamic Traversal}, and \textit{Dynamic Frontier} PageRank with batch updates, consisting purely of edge deletions, increasing from $10^{-7} |E|$ to $0.1 |E|$, in multiples of $10$ (logarithmic scale). The figure on the right illustrates the runtime of each approach for individual graphs in the dataset, while the figure of the left presents overall runtimes (using geometric mean for consistent scaling across graphs).}
  \label{fig:deletions-runtime}
\end{figure*}

\begin{figure*}[hbtp]
  \centering
  \subfigure[Overall result]{
    \label{fig:deletions-speedup--mean}
    \includegraphics[width=0.38\linewidth]{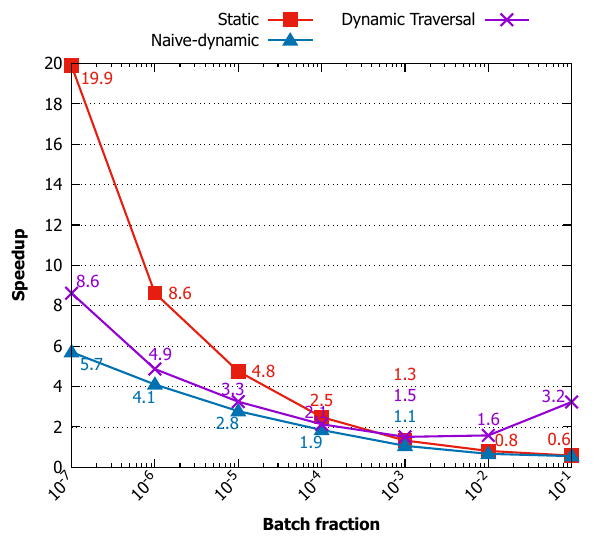}
  }
  \subfigure[Results on each graph]{
    \label{fig:deletions-speedup--all}
    \includegraphics[width=0.58\linewidth]{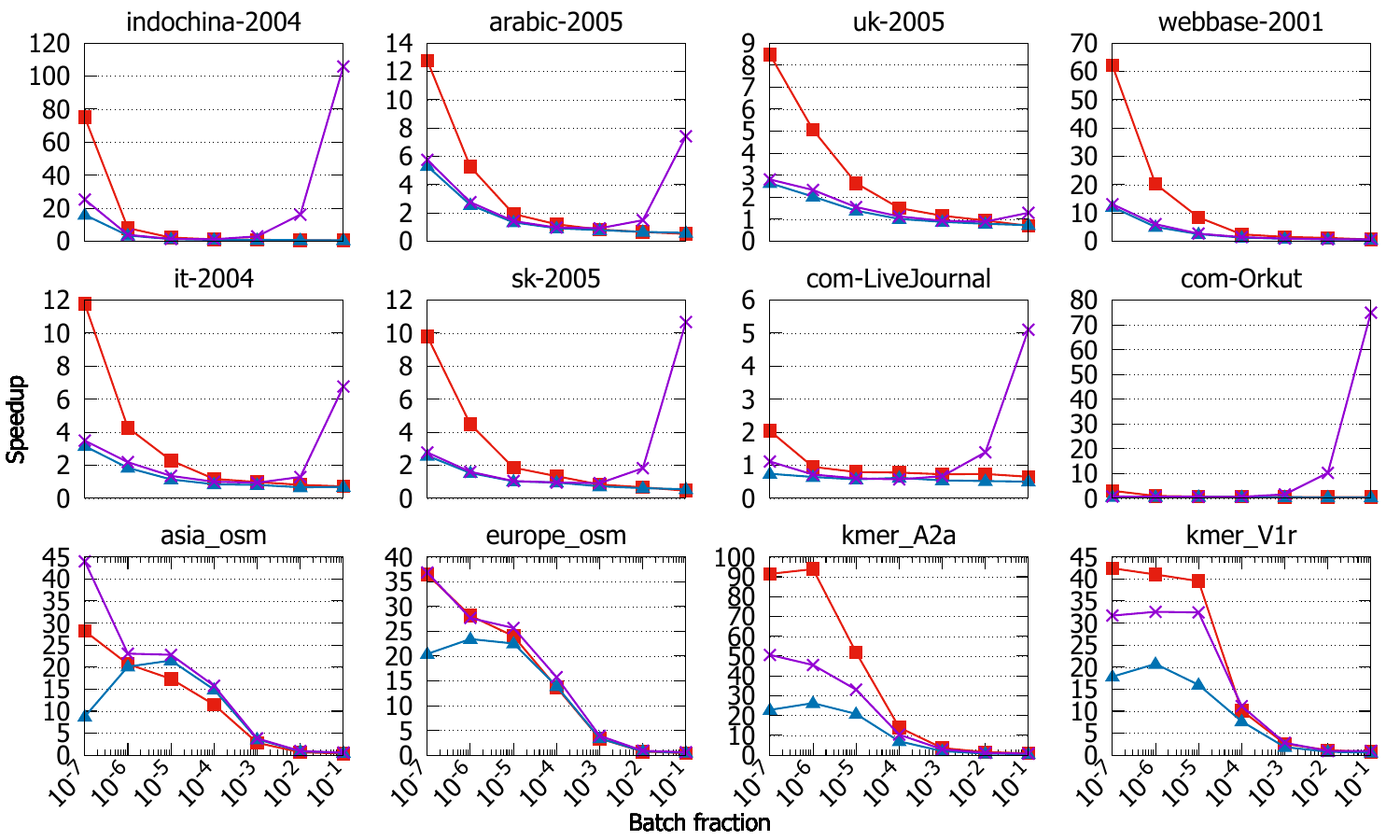}
  } \\[-1ex]
  \caption{Speedup of \textit{Dynamic Frontier} PageRank in relation to \textit{Static}, \textit{Naive-dynamic}, and \textit{Dynamic Traversal} PageRank, on batch updates comprised solely of edge deletions ranging from $10^{-7} |E|$ to $0.1 |E|$ (logarithmic scale). The right figure illustrates the speedup of \textit{Dynamic Frontier} PageRank concerning each approach for individual graphs in the dataset, while the left figure emphasizes the overall speedup.}
  \label{fig:deletions-speedup}
\end{figure*}

\begin{figure*}[hbtp]
  \centering
  \subfigure[Overall result]{
    \label{fig:deletions-error--mean}
    \includegraphics[width=0.38\linewidth]{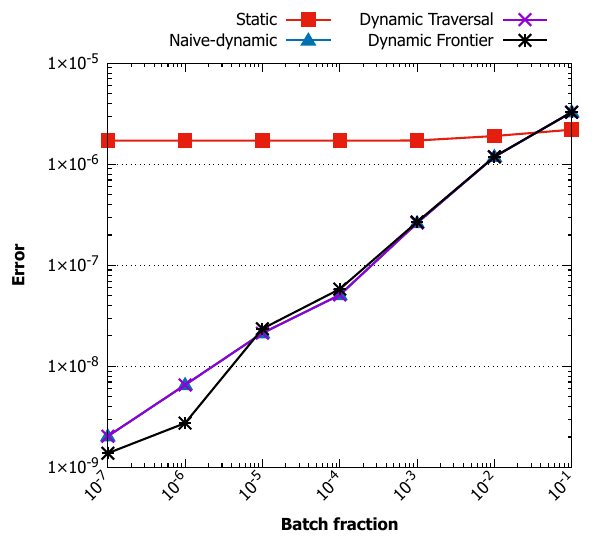}
  }
  \subfigure[Results on each graph]{
    \label{fig:deletions-error--all}
    \includegraphics[width=0.58\linewidth]{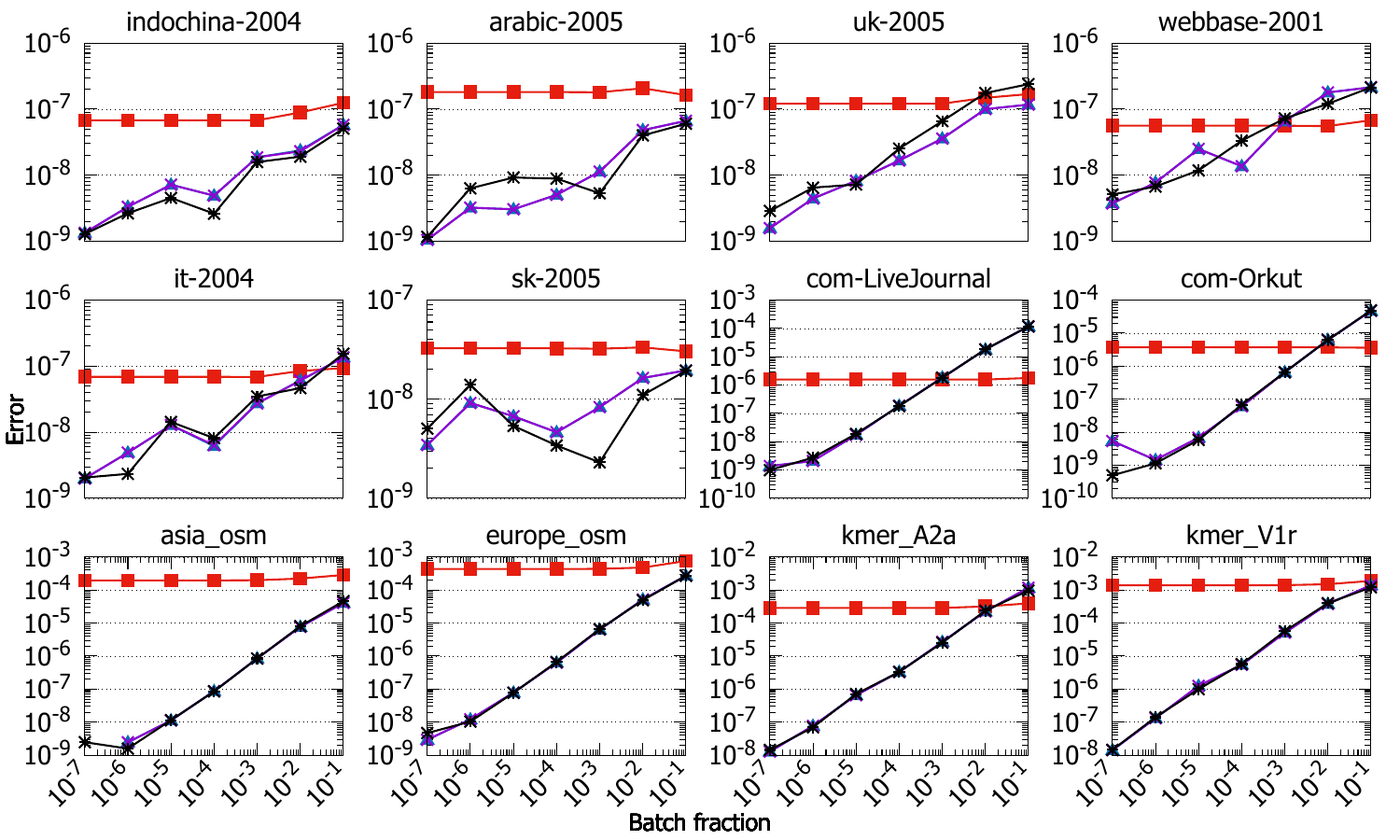}
  } \\[-1ex]
  \caption{Error analysis comparing \textit{Static}, \textit{Naive-dynamic}, \textit{Dynamic Traversal}, and \textit{Dynamic Frontier} PageRank with respect to a Reference Static PageRank (with a tolerance $\tau$ of $10^{-100}$ and limited to $500$ iterations) using $L1$-norm. Batch updates, featuring edge deletions, vary from $10^{-7} |E|$ to $0.1 |E|$ (logarithmic scale). The right figure illustrates the error specific to each approach for individual graphs in the dataset, while the left figure presents overall errors using the geometric mean for consistent scaling across graphs.}
  \label{fig:deletions-error}
\end{figure*}

\begin{figure*}[hbtp]
  \centering
  \subfigure[Overall result]{
    \label{fig:8020-runtime--mean}
    \includegraphics[width=0.38\linewidth]{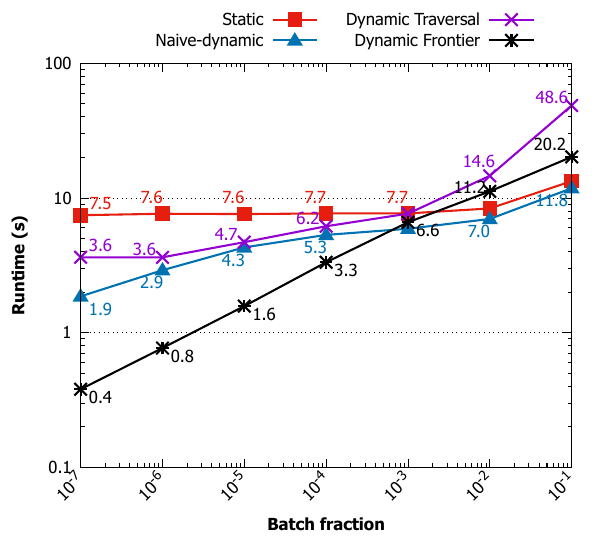}
  }
  \subfigure[Results on each graph]{
    \label{fig:8020-runtime--all}
    \includegraphics[width=0.58\linewidth]{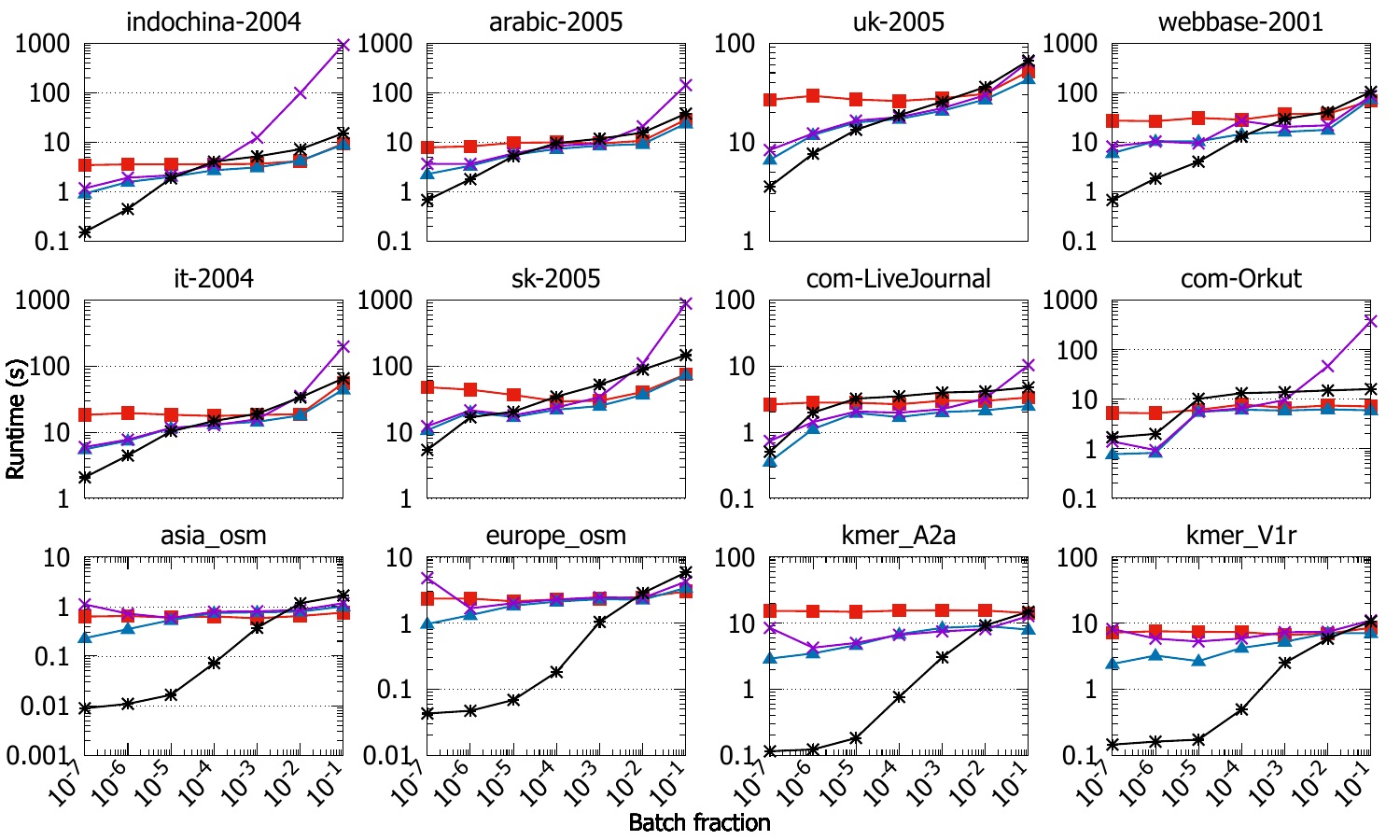}
  } \\[-1ex]
  \caption{Runtime (logarithmic scale) of \textit{Static}, \textit{Naive-dynamic}, \textit{Dynamic Traversal}, and \textit{Dynamic Frontier} PageRank with batch updates increasing from $10^{-7} |E|$ to $0.1 |E|$, in multiples of $10$ (logarithmic scale). The updates include $80\%$ edge insertions and $20\%$ edge deletions, simulating realistic changes upon a dynamic graph. The figure on the right illustrates the runtime of each approach for each graph in the dataset, while the figure of the left presents overall runtimes (using geometric mean for consistent scaling across graphs).}
  \label{fig:8020-runtime}
\end{figure*}

\begin{figure*}[hbtp]
  \centering
  \subfigure[Overall result]{
    \label{fig:8020-speedup--mean}
    \includegraphics[width=0.38\linewidth]{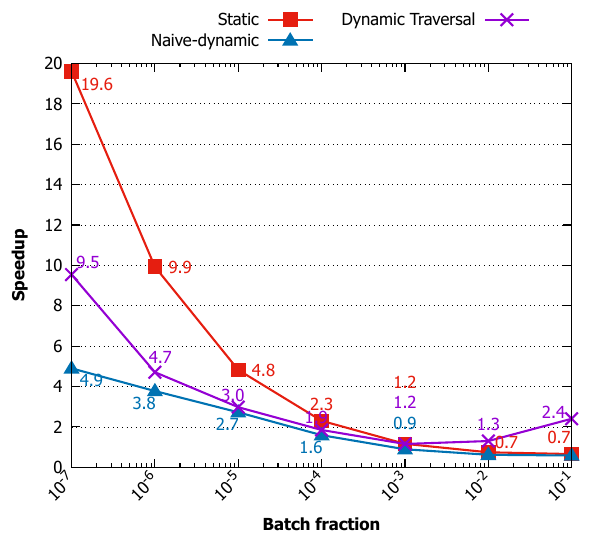}
  }
  \subfigure[Results on each graph]{
    \label{fig:8020-speedup--all}
    \includegraphics[width=0.58\linewidth]{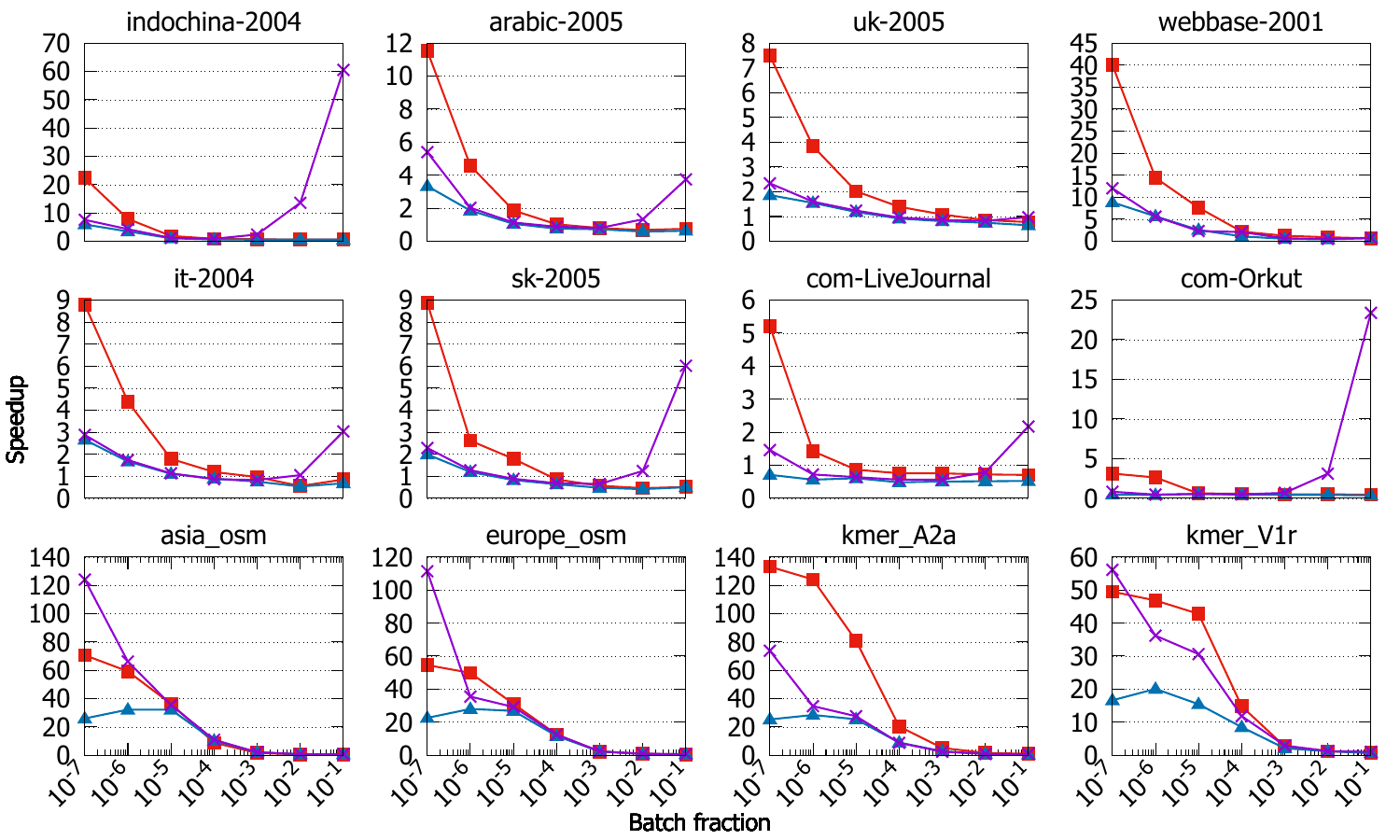}
  } \\[-1ex]
  \caption{Speedup of \textit{Dynamic Frontier} PageRank with respect to \textit{Static}, \textit{Naive-dynamic}, and \textit{Dynamic Traversal} PageRank on batch updates of size $10^{-7} |E|$ to $0.1 |E|$ (logarithmic scale), with $80\%$ edge insertions and $20\%$ edge deletions --- representing a realistic batch update upon a dynamic graph. The figure on the right shows the speedup of \textit{Dynamic Frontier} PageRank, with respect to each approach, for each graph in the dataset --- while the figure of the left highlights the overall speedup.}
  \label{fig:8020-speedup}
\end{figure*}

\begin{figure*}[hbtp]
  \centering
  \subfigure[Overall result]{
    \label{fig:8020-error--mean}
    \includegraphics[width=0.38\linewidth]{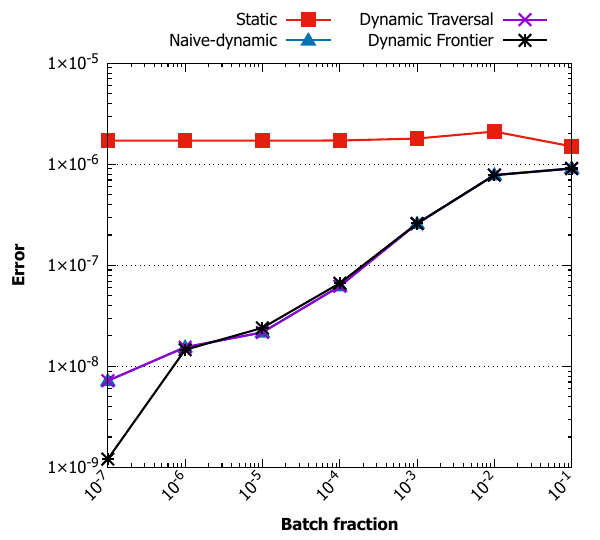}
  }
  \subfigure[Results on each graph]{
    \label{fig:8020-error--all}
    \includegraphics[width=0.58\linewidth]{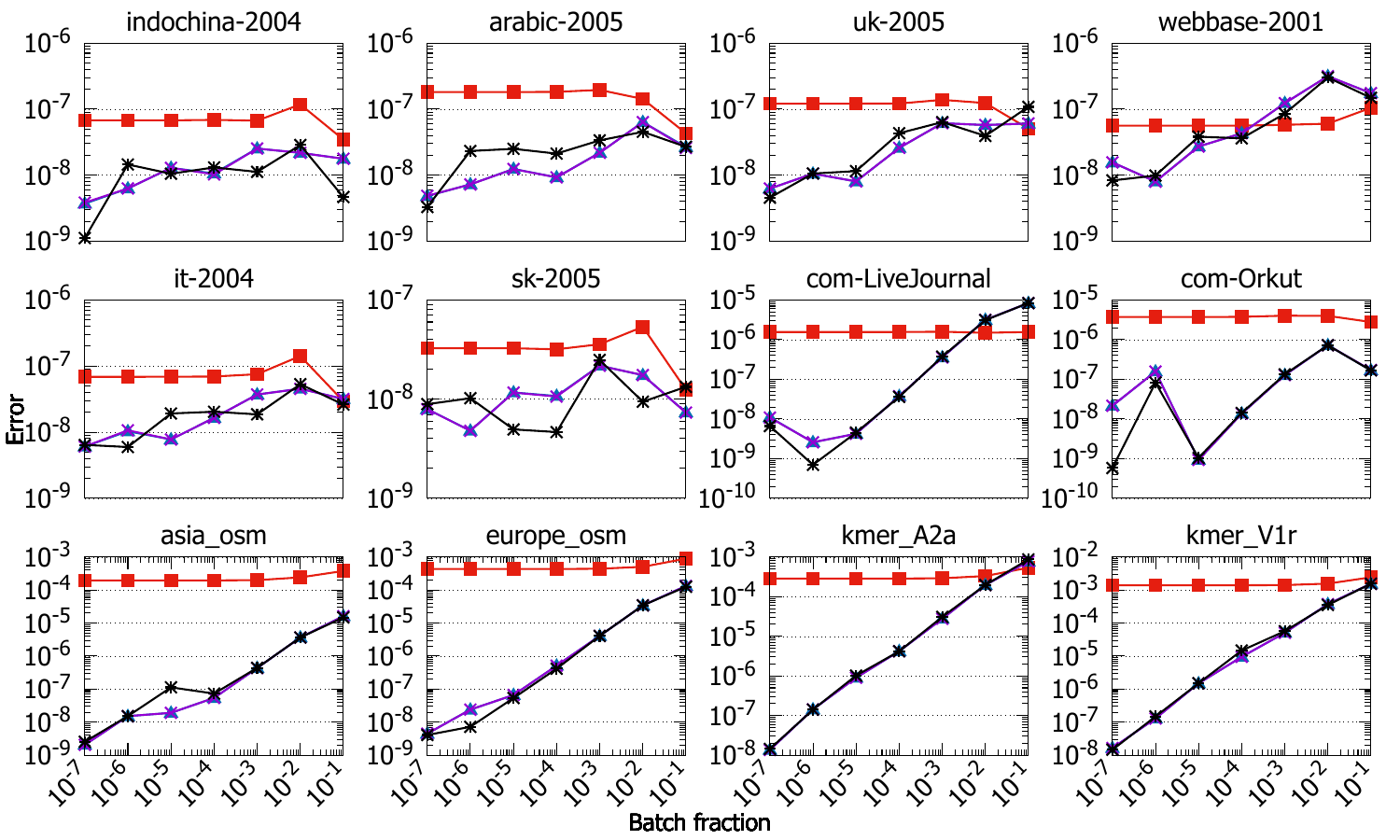}
  } \\[-1ex]
  \caption{Error comparison of \textit{Static}, \textit{Naive-dynamic}, \textit{Dynamic Traversal}, and \textit{Dynamic Frontier} PageRank with respect to a Reference Static PageRank (with a tolerance $\tau$ of $10^{-100}$ and limited to $500$ iterations), using $L1$-norm. Batch updates range from $10^{-7} |E|$ to $0.1 |E|$ (logarithmic scale), consisting of $80\%$ edge insertions and $20\%$ edge deletions to simulate realistic dynamic graph updates. The right figure depicts the error for each approach in relation to each graph, while the left figure showcases overall errors using geometric mean for consistent scaling across graphs.}
  \label{fig:8020-error}
\end{figure*}

\begin{figure}[!hbt]
  \centering
  \subfigure{
    \label{fig:measure-affected--batch}
    \includegraphics[width=0.98\linewidth]{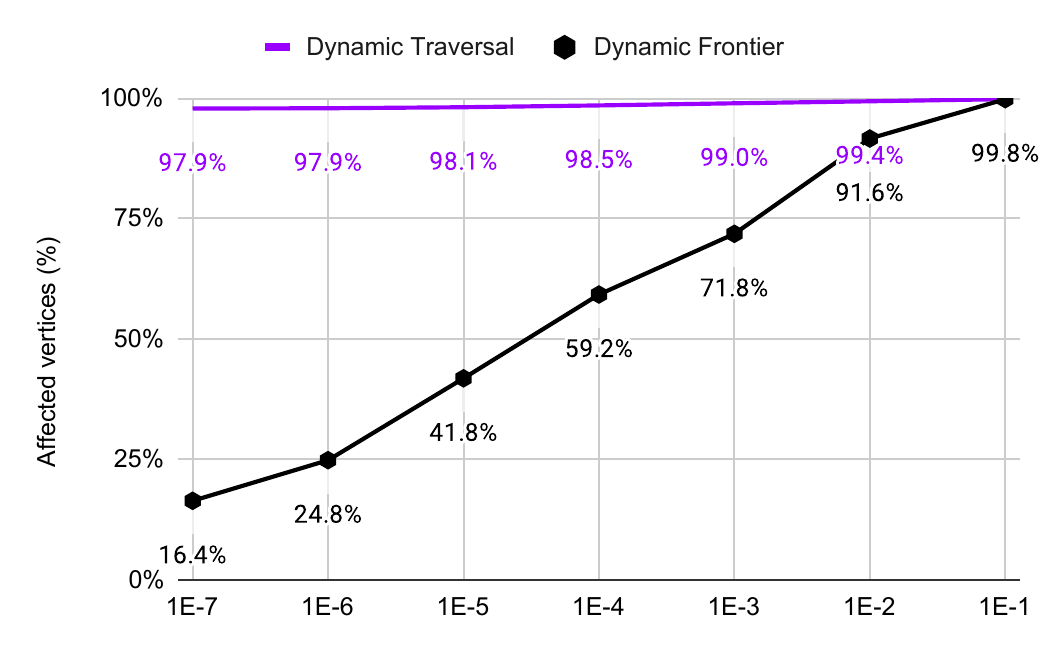}
  } \\[-2ex]
\caption{Average percentage of vertices marked as affected by \textit{Dynamic Traversal} and \textit{Dynamic Frontier} PageRank, with batch size increasing from $10^{-7} |E|$ to $0.1 |E|$ in multiples of $10$ (logarithmic scale), consisting purely of edge insertions. The \textit{Dynamic Frontier} approach marks affected vertices incrementally --- thus, the final percentage (at the end of all iterations) is depicted here.}
  \label{fig:measure-affected}
\end{figure}

\subsection{Performance of Dynamic Frontier PageRank}

We first study the performance of Dynamic Frontier PageRank on batch updates of size $10^{-7}|E|$ to $0.1|E|$ (in multiples of $10$), consisting purely of edge insertions, and compare it with Static, Naive-dynamic, and Dynamic Traversal PageRank. As mentioned above, the edge insertions are generated uniformly at random. Figure \ref{fig:insertions-runtime} plots the runtime of Static, Naive-dynamic, Dynamic Traversal, and Dynamic Frontier PageRank; Figure \ref{fig:insertions-speedup} plots the speedup of Dynamic Frontier PageRank with respect to Static, Naive-dynamic, and Dynamic Traversal PageRank; and Figure \ref{fig:insertions-error} plots the error in ranks obtained with Static, Naive-dynamic, Dynamic Traversal, and Dynamic Frontier PageRank with respect to ranks obtained from a reference Static PageRank (see Section \ref{sec:measurement}). In a similar manner, Figures \ref{fig:deletions-runtime}, \ref{fig:deletions-speedup}, and \ref{fig:deletions-error} present the runtime, speedup, and rank errors of each approach on batch updates consisting purely of edge deletions. Finally, Figures \ref{fig:8020-runtime}, \ref{fig:8020-speedup}, and \ref{fig:8020-error} present the runtime, speedup, and error with each approach on batch updates consisting of an $80\%$ / $20\%$ mix of edge insertions and deletions, in order to simulate realistic batch updates.

\subsubsection{Results with insertions-only batch updates}

Dynamic Frontier PageRank is on average $8.3\times$, $2.7\times$, and $3.4\times$ faster than Static, Naive-dynamic, and Dynamic Traversal PageRank on insertions-only batch updates of size $10^{-7}|E|$ to $10^{-3}|E|$, while obtaining ranks of better accuracy/error than Static PageRank, and of similar accuracy/error as Naive-dynamic and Dynamic Traversal PageRank. On road networks, and protein k-mer graphs, Dynamic Frontier PageRank is significantly faster than its competitors (Naive-dynamic and Dynamic Traversal PageRank).

\subsubsection{Results with deletions-only batch updates}

On deletions-only batch updates of size $10^{-7}|E|$ to $10^{-3}|E|$, Dynamic Frontier PageRank is on average $7.4\times$, $3.1\times$, and $4.1\times$ faster than Static, Naive-dynamic, and Dynamic Traversal PageRank, while obtaining ranks of better accuracy/error than Static PageRank (for batch sizes less than $0.1|E|$), and of similar accuracy/error as Naive-dynamic and Dynamic Traversal PageRank. On \textit{indochina-2004}, \textit{webbase-2001}, road networks, and protein k-mer graphs, Dynamic Frontier PageRank is significantly faster than its competitors (Naive-dynamic and Dynamic Traversal PageRank).

\subsubsection{Results with 80\%-20\% mix batch updates}

On batch updates of size $10^{-7}|E|$ to $10^{-3}|E|$, consisting of $80\%$ insertions and $20\%$ deletions, Dynamic Frontier PageRank is on average $7.6\times$, $2.8\times$, and $4.1\times$ faster than Static, Naive-dynamic, and Dynamic Traversal PageRank, while obtaining ranks of better accuracy/error than Static PageRank, and of similar accuracy/error as Naive-dynamic and Dynamic Traversal PageRank. Similar to deletions-only batch updates, Dynamic Frontier PageRank outperforms its competitors (Naive-dynamic and Dynamic Traversal PageRank) on \textit{indochina-2004}, \textit{webbase-2001}, road networks, and protein k-mer graphs.

\subsubsection{Results with temporal graphs}

We also attempt Static, Naive-dynamic, Dynamic Traversal, and Dynamic Frontier PageRank on temporal graphs found in the Stanford Large Network Dataset Collection \cite{snap14}. On some temporal graphs, Dynamic Frontier PageRank does not outperform its competitors with a frontier tolerance of $\tau_f = \tau / 10^5$, where $\tau$ is the iteration tolerance. However, choosing a lower $\tau_f$ of $\tau / 10$ or $\tau / 100$ allows it achieve good performance. Thus, the choice of frontier tolerance $\tau_f$, possibly in addition to how the frontier of affected vertices is expanded, is dependent upon the nature of the batch update. We plan to explore this in the future.

\subsubsection{Comparison of vertices marked as affected}

Figure \ref{fig:measure-affected} shows the total number of vertices marked as affected (average) by Dynamic Traversal and Dynamic Frontier PageRank on batch updates of size $10^{-7}|E|$ to $0.1|E|$, consisting exclusively of edge insertions. The Dynamic Frontier approach marks affected vertices incrementally --- thus, the final percentage (at the end of all iterations) is depicted in the figure. It is observed that Dynamic Traversal PageRank marks a higher percentage of vertices as affected, even for small batch updates.\ignore{This is likely due the randomly generated edges in the batch update being part of large Strongly Connected Components (SCCs), or due to a large number of such SCCs being reachable from the vertices that are part of the batch update.} In contrast, Dynamic Frontier PageRank marks far fewer vertices as affected, as it incrementally expands the affected region of the graph only after the rank of an affected vertex changes by a substantial amount, i.e., by frontier tolerance $\tau_f = \tau / 10^5$, where $\tau$ is the iteration tolerance (using $L\infty$-norm). In addition, as Dynamic Frontier PageRank incrementally marks vertices as affected, the actual work performed by the algorithm is lower than that indicated by the percentage of affected vertices in Figure \ref{fig:measure-affected}.

\begin{figure}[!hbt]
  \centering
  \subfigure{
    \label{fig:strong-scaling--speedup}
    \includegraphics[width=0.98\linewidth]{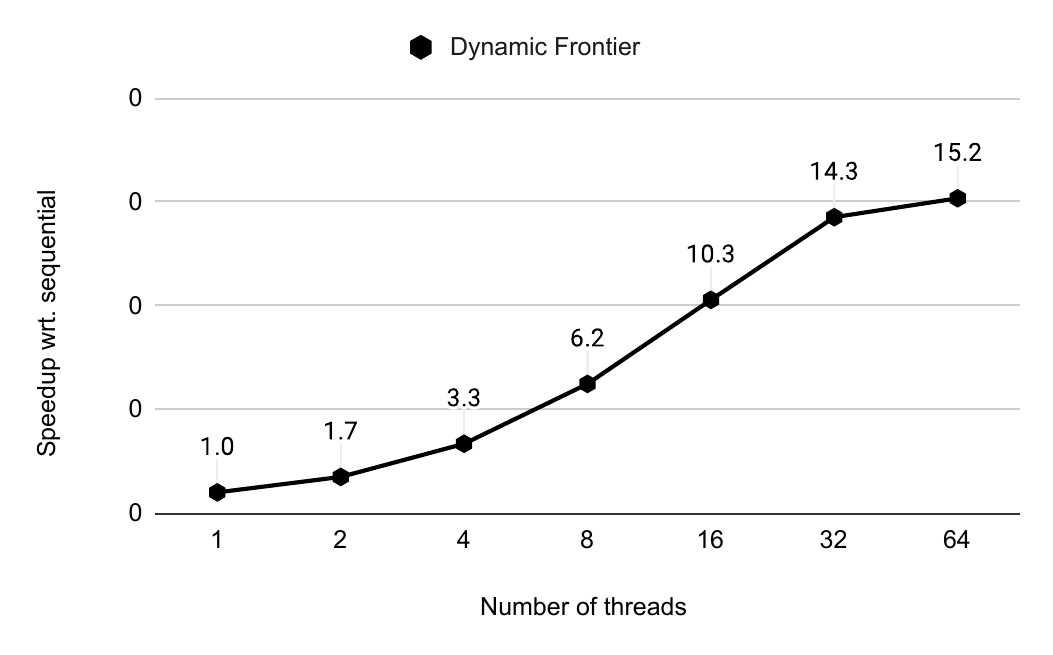}
  } \\[-2ex]
  \caption{Average speedup of \textit{Dynamic Frontier} PageRank with increasing number of threads (in multiples of $2$), on a batch size of $10^{-4}|E|$ (consisting purely of edge insertions).}
  \label{fig:strong-scaling}
\end{figure}

\subsection{Strong Scaling of Dynamic Frontier PageRank}

Finally, we study the strong-scaling behavior of Dynamic Frontier PageRank on batch updates of a fixed size of $10^{-4} |E|$, consisting purely of edge insertions. Here, we measure the speedup of Dynamic Frontier PageRank with an increasing number of threads from $1$ to $64$ in multiples of $2$ with respect to a single-threaded execution of the algorithm. This is repeated for each graph in the dataset, and the results are averaged (using geometric mean).

The results are shown in Figure \ref{fig:strong-scaling}. With $16$ threads, Dynamic Frontier PageRank achieves an average speedup of $10.3\times$, compared to a single-threaded execution, indicating a performance increase of $1.8\times$ for every doubling of threads. At $32$ and $64$ threads, Dynamic Frontier PageRank is affected by NUMA effects (the $64$-core processor we use has $4$ NUMA domains), resulting in a speedup of only $14.3\times$ and $15.2\times$ respectively.

\ignore{\begin{figure}[!hbt]
  \centering
  \subfigure{
    \label{fig:weak-scaling--speedup}
    \includegraphics[width=0.98\linewidth]{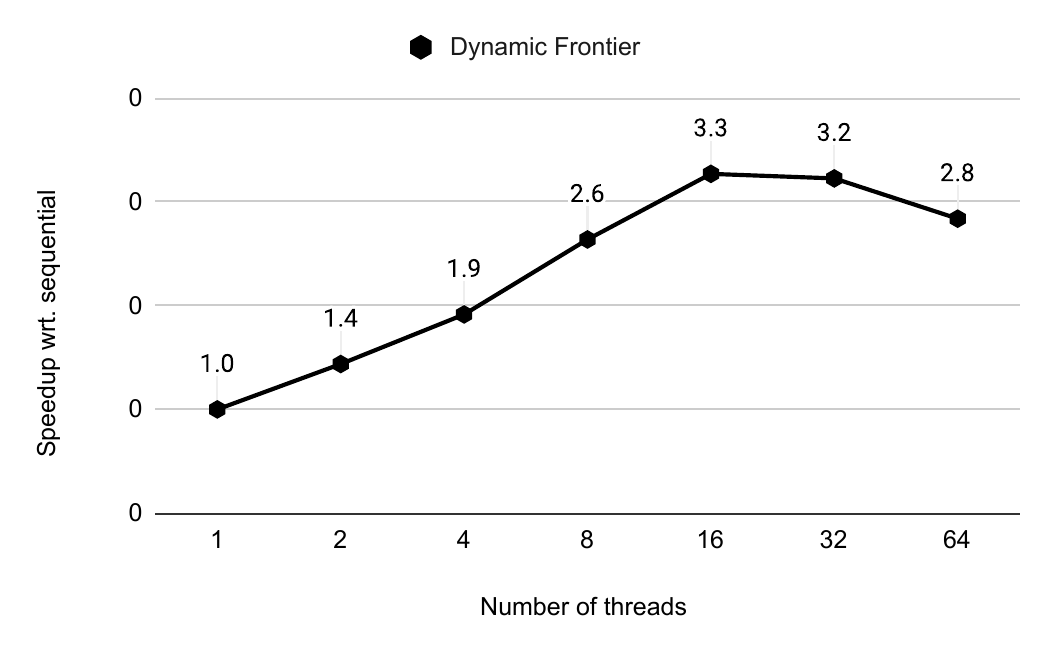}
  } \\[-2ex]
  \caption{Average speedup of \textit{Dynamic Frontier} PageRank with increasing number of threads (in multiples of $2$), on a batch sizes of $10^{-4}|E|$ to $6.4\times10^{-3}|E|$ (consisting purely of edge insertions), increasing in multiples of $2$ in tandem with the increase in the number of threads.}
  \label{fig:weak-scaling}
\end{figure}
}

\section{Conclusion}
\label{sec:conclusion}
In conclusion, this study presents an efficient algorithm for updating PageRank on dynamic graphs. Given a batch update of edge insertions and deletions, our Dynamic Frontier approach identifies an initial set of affected vertices and incrementally expands this set through iterations. On a server with a 64-core AMD EPYC-7742 processor, Dynamic Frontier PageRank outperforms Static, Naive-dynamic, and Dynamic Traversal PageRank by $8.3\times$, $2.7\times$, and $3.4\times$ respectively for uniformly random batch updates of size $10^{-7}|E|$ to $10^{-3}|E|$ with purely edge insertions; $7.4\times$, $3.1\times$, and $4.1\times$ respectively for purely edge deletion updates; and $7.6\times$, $2.8\times$, and $4.1\times$ for updates consisting of an $80\%$ - $20\%$ mix of insertions and deletions. Additionally, the approach exhibits a performance improvement of $1.8\times$ for each doubling of threads. On temporal graphs, we observe that lowering $\tau_f$ to $\tau / 10$ or $\tau / 100$ is needed for Dynamic Frontier PageRank to achieve food performance. Thus, a suitable choice of $\tau_f$ and how the frontier of affected vertices expands depend on the batch update's nature. We plan to explore this in the future.

\begin{acks}
I would like to thank Prof. Kishore Kothapalli, Prof. Sathya Peri, and Prof. Hemalatha Eedi for their support.
\end{acks}

\bibliographystyle{ACM-Reference-Format}
\bibliography{main}

\end{document}